\documentclass[review,1p]{elsarticle}

\usepackage{comment}

\usepackage{graphicx} 

\usepackage{multirow}

\usepackage[dvipsnames]{xcolor}

\usepackage{amsmath}
\usepackage[none]{hyphenat}
\usepackage{amsfonts}
\usepackage{amssymb}
\usepackage{amsthm}

\usepackage{float}

\usepackage{gensymb}

\let\oldsqrt\sqrt
\def\sqrt{\mathpalette\DHLhksqrt}
\def\DHLhksqrt#1#2{%
	\setbox0=\hbox{$#1\oldsqrt{#2\,}$}\dimen0=\ht0
	\advance\dimen0-0.2\ht0
	\setbox2=\hbox{\vrule height\ht0 depth -\dimen0}%
	{\box0\lower0.4pt\box2}}

\usepackage{siunitx}
\usepackage{amsmath}
\usepackage{amsfonts}
\usepackage{amssymb}
\usepackage{amsthm}

\usepackage{float}
\usepackage{braket}
\usepackage{booktabs} 

\usepackage{placeins}

\usepackage{adjustbox}

\usepackage[utf8]{inputenc}
\usepackage{hyperref}
\usepackage{subfig}
\usepackage{microtype}
\usepackage{mathtools}
\usepackage{setspace}
\DeclareSIUnit\neutron{neutron}
\DeclareSIUnit\atom{atom}
\DeclareSIUnit\shake{shake}
\DeclareSIUnit\collision{collision}

\usepackage{nomencl}
\makenomenclature

\renewcommand{\nomgroup}[1]{%
      \ifthenelse{\equal{#1}{R}}{%
        \item[\textbf{Roman symbols}]}{%
        \ifthenelse{\equal{#1}{G}}{%
          \item[\textbf{Greek symbols}]}{%
          \ifthenelse{\equal{#1}{S}}{%
            \item[\textbf{Subscripts}]}{%
            \ifthenelse{\equal{#1}{T}}{%
              \item[\textbf{Superscripts}]}{}}}}}%

\begin{document}
\sloppy
	
	\title{Experimental and numerical characterization of single-phase pressure drop and heat transfer enhancement in helical corrugated tubes}
	
    \author[1,2]{Gon\c{c}alo G. Cruz}
    \author[2]{Miguel A.A. Mendes\corref{cor}}
    \ead{miguel.mendes@tecnico.ulisboa.pt}
    \author[2]{José M.C. Pereira}
    \author[3]{H. Santos}
    \author[4]{A. Nikulin}
    \author[1,5]{Ana S. Moita\corref{cor}}
    \ead{anamoita@tecnico.ulisboa.pt}

   \address[1]{IN+ – Center for Innovation, Technology and Policy Research, Instituto Superior Técnico, Universidade de Lisboa, Av. Rovisco Pais, 1049-001 Lisboa, Portugal}
    \address[2]{LAETA, IDMEC, Instituto Superior Técnico, Universidade de Lisboa, Av. Rovisco Pais 1, 1049-001 Lisboa, Portugal}
    \address[3]{ADAI-LAETA, IPLeiria Delegation, School of Technology and Management, Polytechnic Institute of Leiria, Leiria, Portugal}
    \address[4]{Centre for Cooperative Research on Alternative Energies (CIC energiGUNE), Basque Research and Technology Alliance (BRTA), Alava Technology Park, Albert Einstein 48, 01510 Vitoria-Gasteiz, Spain}
    \address[5]{CINAMIL - Center for Investigation, Development and Innovation of the Military Academy, Department of Exact Sciences and Engineering, Portuguese Military Academy, R. Gomes Freire, 203, 1169-203 Lisboa, Portugal}

\cortext[cor]{Corresponding authors}
	
\begin{abstract}

The internal flow in corrugated tubes of different helical pitch, covering from the laminar to turbulent regime, was studied in order to characterize the three-dimensional flow and the influence of corrugation geometry on pressure drop and convective heat transfer.  With water as working fluid and an imposed wall heat flux, ranging from around 4 to 33~kW/m$^2$, a numerical model was developed with a CFD commercial software, where $k-\omega$ SST was used to model turbulence. Experimental tests were performed covering Reynolds numbers in the range from around 300 up to 5000, which allowed to identify the transition region and validate the numerical model. The results show that due to the swirl induced by the corrugation, the critical Reynolds number for the start of transition to turbulent flow is reduced. The thermal performance factor, which quantifies the heat transfer enhancement at the expense of pressure drop, was used to compare the corrugated tubes against the reference case of smooth tube. Based on this, all investigated corrugated geometries performed better than the smooth tube, except for low Reynolds numbers (${\rm Re}<500$) in the laminar regime. Overall, the corrugated tube with the lowest pitch showed a clearly better performance for the intermediate range of Reynolds numbers ($1000<{\rm Re}<2300$), being an optimal choice for a wider range of operating conditions in the transitional and turbulent flow regimes. 


\end{abstract}
	\maketitle



\section{Introduction}
\label{sec:intro}

\begin{figure}[H]
  \centering
  \includegraphics[width=0.5\textwidth]{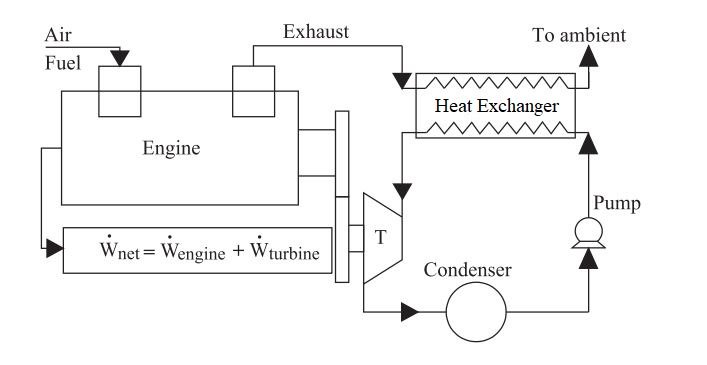}
  \caption{Schematic representation of an engine waste heat recovery Rankine cycle system; adapted from Wang et al.~\cite{Wang2011}}.
  \label{fig:Basic EHR scheme}
\end{figure}

Energy efficiency is one of the main economical and technological challenges of the century, mainly due to sustainability and environmental concerns. At the industrial level, the coupling of the power generation gas turbine cycle with a Rankine cycle, through a heat recovery steam generator, has been an important step targeting energy efficiency increase~\cite{Franco2002}. Nevertheless, the concept of heat recovery is also emerging in transportation systems, with the utilization of the exhaust gases waste heat from internal combustion (IC) engines (typically at temperatures between 500 and 650$\degree$C in heavy-duty vehicles). In this respect, various possibilities of waste heat recovery systems are reported and reviewed in the literature~\cite{Wang2011}. From the different alternative approaches, combination of IC engine with a Rankine cycle, as represented in Figure~\ref{fig:Basic EHR scheme}, has a high potential in vehicle applications, which allows the increase of combined cycle efficiency using energy that otherwise would be dissipated as waste heat to the environment. This concept has a large potential when applied in heavy-duty vehicles for both on-road and off-road applications, where the system compactness is mandatory. Water, ethanol or refrigerant fluids (like R245fa) are usually chosen as working fluid for the Rankine cycle, where a major advantage of water is the fact that it is safe to use.

Let us consider, for example, a cross flow tube heat exchanger (CFT-HEX) for a vehicle Rankine cycle waste heat recovery system.  A CFT-HEX consists of a tube bank usually arranged in an inline or staggered manner, where exhaust gas flows across the tubes and the working fluid circulates inside the tubes, allowing these heat exchangers to be designed for high pressures. On the working fluid side, the heat exchanger (HEX) has three zones: the pre-heater (liquid), the evaporator (two-phase) and the super-heater (vapour). Considering the exhaust gas and the working fluid operating conditions, the pre-heater section corresponds to the largest part of the HEX, accounting for about 70\% of the HEX volume. Furthermore, the power density (thermal power per unit of volume) on the pre-heater section of the HEX is very low as compared with that on the evaporator and super-heater sections. Accordingly, a more compact HEX strongly depends on the increase of the power density at the pre-heater section. Due to the low mass flow rates of the working fluid, the Reynolds number for a smooth tube leads to a laminar flow regime, which greatly limits the heat transfer rates.

The efficiency of the Rankine cycle strongly depends on the efficiency of the HEX itself (see Figure~\ref{fig:Basic EHR scheme}). In this respect, one must account for both the heat transfer rate as well as the pumping power requirements for circulating the fluid streams, where the balance between these two factors allows to define the thermal performance of a HEX. Additionally, another important characteristic of a HEX is its compactness, \textit{i.e.} the HEX volume per unit of heat transfer rate, which is particularly important for non-stationary applications (such as transportation systems).

Various heat transfer enhancement methods have been discussed in the literature, with the goal of increasing the HEX compactness and thermal performance; see \textit{e.g.}~\cite{Sheikholeslami2015}. These techniques can be classified as active or passive, depending on whether they require or not external power. Besides the obvious advantage of not requiring an external power source, passive techniques are further valuable as they are cost effective and easy to implement.

Passive heat enhancement techniques typically use secondary flows and unsteadiness to increase heat transfer, mainly by geometrical modifications (\textit{e.g.} dimpled or corrugated tubes) and extended surfaces (\textit{e.g.} smooth tubes with fins, twisted tapes or wire-coil inserts); see \textit{e.g.}~\cite{Ji2015}. Hybrid passive techniques can also be used, such as the coupling of modified geometries with the addition of nanoparticles to the original working fluid, see e.g.~\cite{WEI2021100948,WANG201951}. In this respect, by adding 0.5~\%~wt. TiO$_2$ nanoparticles to water, the heat transfer performance in corrugated tubes was found to be enhanced by 4.8–66.3~\% compared with pure water~\cite{WANG201951}.

Garc\'ia et al.~\cite{Garcia2012} carried out a comparative study between corrugated, dimpled and wire-coil-inserted tubes, where they concluded that these modified tube geometries have a greater influence on the pressure drop characteristics than on the heat transfer enhancement. Associated to this, the induced artificial roughness was found to strongly affect the advance of transition to turbulence (when compared to a smooth tube). For Reynolds numbers at the end of the laminar-turbulent transition regime, corrugated tubes presented the lowest friction factor, while the Nusselt number remained similar for all modified tube geometries. 

\begin{figure}[!htb]
    \centering
    \includegraphics[width=0.5\textwidth]{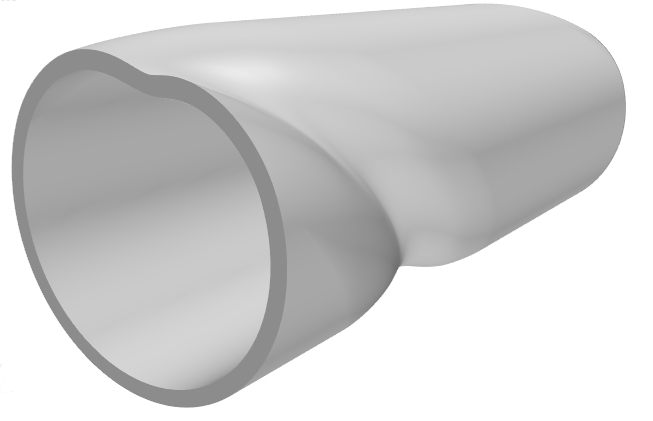}
    \caption{CAD model view of an helical corrugated tube (P12).}
    \label{fig:corrugated real}
\end{figure}

In the extensive review work from Kareem et al.~\cite{Kareem2015} it is mentioned that of all the different corrugations, the helical corrugated tubes provide higher heat transfer enhancement. They concluded that this type of geometry (see Figure~\ref{fig:corrugated real}), in addition to providing a larger surface contact area, also induces a swirl effect on the flow that is the main contributor to the heat transfer enhancement.

From a more practical perspective, the use of helical corrugated tubes for HEX applications also has some advantages, namely their production costs are similar to that of a smooth tube and their fouling removal is relatively easy when compared to other modified tube geometries~\cite{Ji2015}. In this respect, Qian etal.~\cite{QIAN2020119876} analysed the thermo-hydraulic performance of multi-start spirally corrugate tubes, while combining performance evaluation criteria and entropy production theorem with the help of numerical simulations, which is a useful tool for further heat transfer augmentation of HEXs.

Akbarzadeh et al.~\cite{AKBARZADEH2020735,AKBARZADEH2021101013} investigated the potential of applying a helical corrugated absorber tube in a parabolic trough collector (PTC), where the effect of different different pitch ratios and rib-heights ratios on the performance of the PTC was assessed. They concluded that corrugated tubes are very suitable for this type of application and present the highest performance in the transitional regime.

In summary, helical corrugated tubes appear as an appealing solution for compact HEXs, which can be used for example in exhaust heat recovery systems implemented on heavy-duty vehicles~\cite{Liu2013}. Although there are multiple corrugation geometries available~\cite{Kareem2015,Wang2018b}, they all share the common feature of increasing the heat transfer at the expense of a higher pressure drop, mostly caused by the swirl induced in the flow. In this context, the transitional flow regime seems to allow a good compromise between low pressure drop and high heat transfer~\cite{Garcia2012,Meyer_2014}.

As an extension of the previous experimental study from Andrade et al.~\cite{ANDRADE2019940}, the main objective of the present work is to better characterize the influence of different helical corrugation geometries on single-phase pressure drop and heat transfer enhancement. Therefore, the internal flow in corrugated tubes of different helical pitch was investigated, covering the transition from laminar to turbulent regime, while coupling experimental results with detailed 3D model predictions in order to investigate and describe the main features of the fluid flow and heat transfer processes.

Indeed, the experimental and numerical results obtained for the corrugated tubes investigated in this study will be used to further optimize a CFT-HEX for a vehicle Rankine cycle waste heat recovery application. For such an application, HEX compactness is of crucial importance. As compared to our previous research~\cite{ANDRADE2019940}, the internal tube diameter of the corrugated tubes was further reduced from the 5.75~mm (originally) to 4.5~mm in the present study.

It is important to point out that experimental and numerical studies on corrugated tubes reported in the literature focus mainly on refrigeration systems applications that use tubes with high internal diameter, typically larger than 14~mm. So, the present study broadens the experimental data available on corrugated tubes under transitional flow regime. Furthermore, the literature survey revealed that numerical investigations on the corrugated tubes were mainly carried out under turbulent flow conditions with Re~$>$~5000; see e.g.~\cite{Wang2018b,Mohammed2013,CORCOLES2020106526,YANG2020106520}. Hence, the present study contributes to close the experimental and numerical research gap on the heat transfer and pressure drop characteristics in the transitional flow regime for corrugated tubes with low internal diameter (4.5~mm). To investigate the entire transitional flow regime, and for continuity purposes, the experimental and numerical results were also given in the laminar and turbulent flow regimes with Reynolds numbers in the range from around 300 up to 5000. Furthermore, to accomplish with the practical applications requisites, the imposed wall heat flux was further increase from 21.1~kW/m$^2$~\cite{ANDRADE2019940} to 33~kW/m$^2$ in the present study. To the best of authors knowledge, this is the maximum heat flux reported in literature studies for the study of corrugated tubes under transitional flow regime.

\section{Heat transfer and pressure drop in helical corrugated tubes}
\label{sec:background}

\begin{figure}[!htb]
    \centering
    \includegraphics[width=0.5\textwidth]{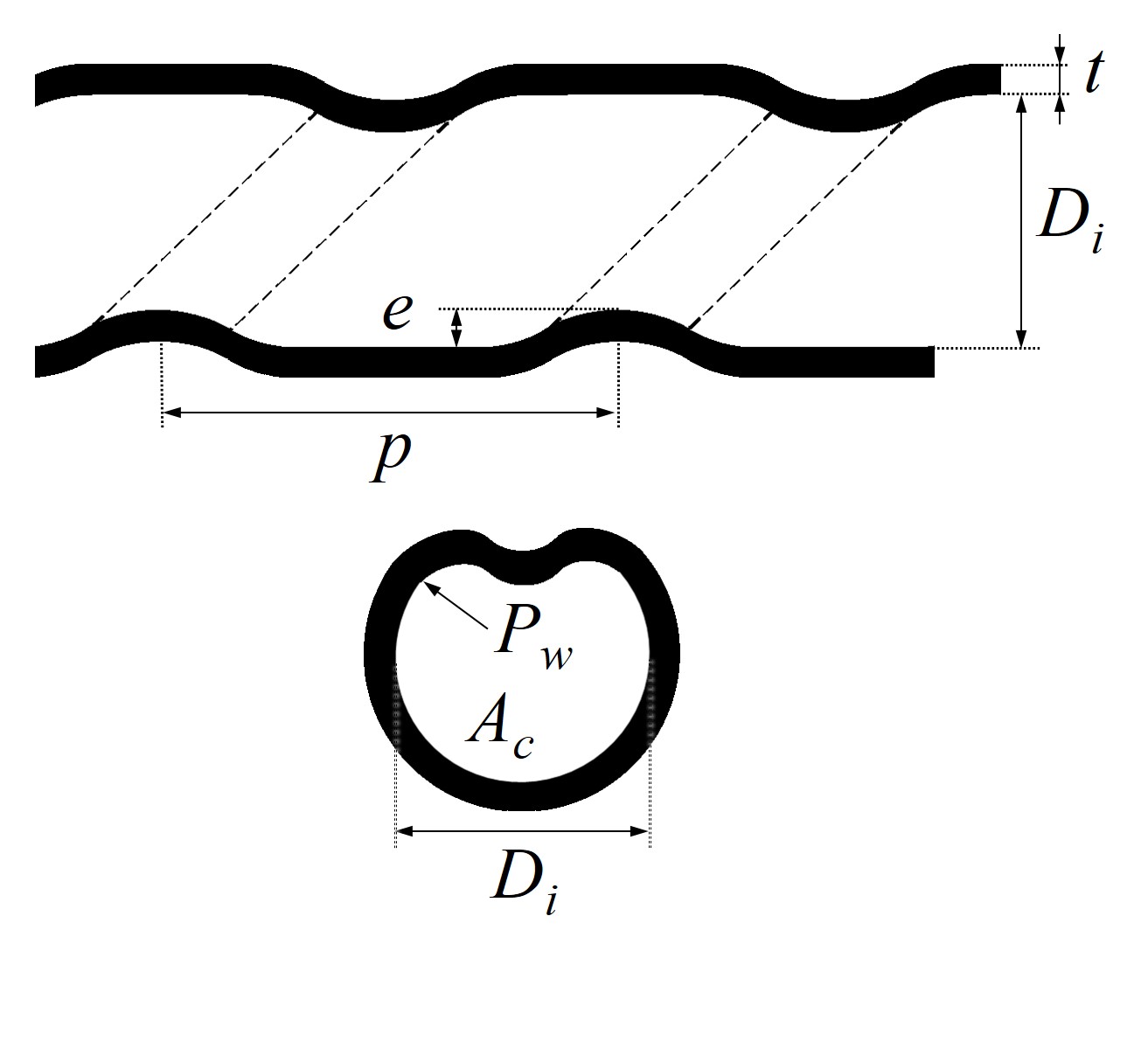}
    \caption{Sketch of an helical corrugated tube with characteristic dimensions, showing views of axial-section (top) and cross-section (bottom).}
    \label{fig:corrugated dimensions}
\end{figure}

Figure~\ref{fig:corrugated dimensions} depicts a sketch of the geometry of an helical corrugated tube with inner diameter $D_{i}$ and wall thickness $t$, where the characteristic corrugation dimensions are shown: the helical pitch $p$ and height $e$. 

Considering the characteristic dimensions of the corrugated geometry shown in Figure~\ref{fig:corrugated dimensions}, the roughness degree of the tube wall can be quantified by evaluating the severity index defined as~\cite{Vicente2004,Vicente2004a}:
     \begin{equation}
     \phi = \frac{e^2}{p D_{i}}
    \label{eq:severity_index}
    \end{equation}
which combines the effects of both helical pitch and height of corrugation.

Moreover, based on the cross-section area $A_{c}$ available for the flow and the associated wetted perimeter $P_{w}$ (see Figure~\ref{fig:corrugated dimensions}), the hydraulic diameter of the tube is given by definition as:
     \begin{equation}
     D_h \equiv \frac{4 A_c}{P_w}
    \label{eq:hydro_diameter}
    \end{equation} 
Due to the corrugation, $A_{c}$ is reduced when compared to a smooth tube of equal inner diameter. However, $P_{w}$ may be increased or slightly reduced depending on the corrugation height and shape. Nevertheless, the hydraulic diameter of the corrugated tube tends to be generally smaller when compared to that of a smooth tube (for which $D_h = D_i$).

Knowing the mass flow rate $\Dot{m}$ inside the tube, the average axial velocity of the flow can be calculated as:

\begin{equation}
u_m = \frac{ \dot{m}}{\rho \, A_c}
\label{eq:u_medio}
\end{equation}

where $\rho$ is the fluid density. Due to the reduction in $A_c$, the corrugated tube presents a higher $u_m$ than the smooth tube (of same $D_i$) under similar flow conditions.

Based on the fluid properties and using $D_h$ and $u_m$ as the characteristic length and velocity, respectively, the Reynolds, Nusselt and Prandtl numbers are defined as:
\begin{eqnarray}
\label{eq:reynolds}
{\rm Re} &=& \frac{ \rho \, u_m D_{\rm h}}{\mu}\\
 \label{eq:Nusselt}
{\rm Nu} &=& \frac{h \, D_h}{k}\\
\label{eq:prandtl}
{\rm Pr} &=& \frac{c_p \, \mu}{k}
\end{eqnarray}
where $k$, $\mu$ and $c_p$ are the thermal conductivity, dynamic viscosity and specific heat capacity of the fluid, respectively. Moreover, $h$ is the local convective heat transfer coefficient between the fluid and tube inner wall, used for expressing the Nusselt number.

Instead of ${\rm Re}$ and ${\rm Nu}$, some authors (e.g. Vicente et al. ~\cite{Vicente2004,Vicente2004a}) prefer to calculate Reynolds and Nusselt numbers ($\widetilde{\rm Re}$ and $\widetilde{\rm Nu}$, respectively) using the inner diameter $D_i$ as characteristic length. It is important to highlight that the results presented in this work consider ${\rm Re}$ and ${\rm Nu}$ defined based on the hydraulic diameter $D_h$; see Eqs.~\eqref{eq:reynolds} and~\eqref{eq:Nusselt}. Therefore, the proper comparison with literature results, obtained using $\widetilde{\rm Re}$ and $\widetilde{\rm Nu}$ definitions, can be carried out by simply applying the following conversion relations: $\widetilde{\rm Re} = \left(D_i/D_h \right) {\rm Re}$ and $\widetilde{\rm Nu}= \left(D_i/D_h \right) {\rm Nu}$. It is worth mentioning that for a smooth tube, $D_i$=$D_h$, so that $\widetilde{\rm Re} = {\rm Re}$ and $\widetilde{\rm Nu}= {\rm Nu}$.

Since heat transfer enhancement is generally obtained at the expense of energy dissipated by extra friction, one way of studying the effect of tube corrugation on the HEX performance is to compare the corrugated data results (heat transfer and pressure drop) against the reference case of a smooth tube~\cite{Vicente2004,Dizaji2016,Pethkool2011}. In this respect, one can define a thermal performance factor as~\cite{Bergles_Blumenkrantz_Taborek_1974,ZIMPAROV19941807}:
  \begin{equation}
     \eta = \frac{{{\rm Nu}}/{{\rm Nu}_{ref}}}{({f}/{f_{ref}})^{1/3}}
    \label{eq:thermal performance}
\end{equation} 
where ${\rm Nu}$ and $f$ are the Nusselt number and friction factor, respectively, for the corrugated tube and, ${\rm Nu}_{ref}$ and $f_{ref}$ are the corresponding values for the smooth tube, evaluated at the same Reynolds number value. The definition of $\eta$, given by Eq.\eqref{eq:thermal performance}, imposes equal pumping power for both corrugated and smooth tubes. Therefore, its numerator represents the heat transfer enhancement promoted by the corrugated tube (with respect to the smooth tube), whereas its denominator reflects the associated increase in pressure drop. 

Different correlations of ${\rm Nu}$ and $f$ for corrugated tubes can be found in the literature, which have been obtained based on either experimental~\cite{Vicente2004,Vicente2004a,Pethkool2011} or numerical~\cite{Wang2018b,Wang2018a} data. However, most of these correlations are limited to specific corrugation geometries or they are only valid in certain flow regimes. In this respect, the experimental work of Vicente et al.~\cite{Vicente2004,Vicente2004a} covers both laminar and turbulent flow regimes for ten different corrugated tube geometries and using two different working fluids (water and ethylene glycol), where their proposed correlations consider the definition of Reynolds and Nusselt numbers based on the tube inner diameter $D_{i}$, $\widetilde{\rm Re}$ and $\widetilde{\rm Nu}$.

For the fully developed laminar regime, the Darcy friction factor correlation is given as~\cite{Vicente2004}:
     \begin{equation}
    f = 119.6 \ \phi^{0.11} \; {\widetilde{\rm Re}}^{-0.97}
    \label{eq:vicente_f_lam}
    \end{equation} 
while for the fully developed low Reynolds turbulent regime ($2000<{\widetilde{\rm Re}}<8000$), the following correlation was provided~\cite{Vicente2004a}:
\begin{equation}
    f = 6.12 \ \phi^{0.46} \; {\widetilde{\rm Re}}^{-0.16}
    \label{eq:vicente_f_turb}
\end{equation}
which is recommended for tubes with soft corrugation, \textit{i.e.} when the severity index is $\phi<10^{-3}$. It is worth mentioning that the above correlations are expressed in terms of Darcy friction factor $f$, which is equal to four times the Fanning friction factor originally considered in~\cite{Vicente2004,Vicente2004a}.

The dependency of Eqs.~\eqref{eq:vicente_f_lam} and~\eqref{eq:vicente_f_turb} on the Reynolds number is quite similar to that found in the analogous equations for a smooth tube, \textit{i.e.} the Hagen-Poiseuille and Blasius equations for the fully developed laminar and turbulent flows, respectively~\cite{bejan2013convection}. However, they include an extra term, which adds the dependence on corrugation geometry in the form of the severity index $\phi$.
   
Vicente et al.~\cite{Vicente2004a} also proposed a general correlation for the Nusselt number in the fully developed turbulent flow regime:
    \begin{equation}
    \widetilde{\rm Nu} = 0.374 1 \phi^{0.25} \left[ {\widetilde{\rm Re}} - 1500 \right]^{0.74}{\rm Pr}^{0.44}
    \label{eq:vicente_nu_turb}
    \end{equation}   
which is applicable for ${\widetilde{\rm Re}}>2000$ and also shows a dependency on $\phi$. Based on Eq.~\eqref{eq:vicente_nu_turb}, the authors concluded that the heat transfer enhancement of corrugated tubes is more pronounced at higher Prandtl numbers since, for smooth tubes, they observed a weaker Prandtl number dependency in the form $ \widetilde{\rm Nu} \propto {\rm Pr}^{0.37}$.

In another work from Vicente et al.~\cite{Vicente2004}, a correlation for $\widetilde{\rm Nu}$ was proposed in the laminar regime. However, it was given as function of the Rayleigh number, since the heat transfer experiments were governed by mixed convection (\textit{i.e.} non-negligible buoyancy effect), mainly due to the large diameter of the employed corrugated tubes ($D_i = 18$~mm). Nevertheless, the present work considers corrugated tubes of smaller diameter ($D_i = 4.5$~mm), where the dominant heat transfer mechanism is found to be forced convection (see section~\ref{sec:experimental-data}). Therefore, the present results are outside the applicability range of their correlation.

From previous investigations of pressure drop and heat transfer in both laminar and turbulent flow regimes, the laminar-turbulent transition in corrugated tubes was found to occur in a gradual smooth way~\cite{Garcia2012,Vicente2004}. In this respect, data from Vicente et al.~\cite{Vicente2004} show that the transition point is mainly dependent on the corrugation height $e$. Therefore, they proposed a correlation for estimating the critical Reynolds number of transition, expressed as:
\begin{equation}
\widetilde{\rm Re}_{cr}=2100\left[1+1.18\times10^{7}(e / D_{i})^{3.8}\right]^{-0.1}
\label{eq:crit_re}
\end{equation}
which does not account for the effect of corrugation helical pitch $p$, because this was found to be negligible. Equation~\eqref{eq:crit_re} provides a possibility for operation design, taking into consideration that the desired flow regime (laminar or turbulent) can be simply controlled by the corrugation height $e$. 

Recently, C\'orcoles-Tendero et al.~\cite{Corcoles-Tendero2018a} used the experimental results for corrugated tubes from Vicente et al.~\cite{Vicente2004a} in order to validate 3D model predictions of pressure drop and heat transfer in the turbulent regime (using the realizable k-epsilon model for turbulence). Their validation study indicated that the numerical model was able to predict an average Nusselt number within a maximum relative error of 17\%, when compared with experimental data. Furthermore, the differences in the friction factor were found to be lower than 9\%. Therefore, their study suggests that similar 3D simulations can be used as a complement to heat transfer measurements in corrugated tubes, providing additional detailed information that is generally difficult to access from an experimental approach (\textit{e.g.} local velocity or temperature profiles).

In this context, numerical studies have been focused on the effects of modifying corrugation geometry, specially with the goal of increasing the thermal performance $\eta$ of the heat transfer process; see e.g.~\cite{Wang2018b,CORCOLES2020106526,YANG2020106520}. Wang et al.~\cite{Wang2018b} carried out a sensitivity analysis and an optimization study for helical corrugated tubes in the turbulent regime (using the Reynolds stress transport model for turbulence), where they concluded that the friction factor and Nusselt number are both strongly influenced by the Reynolds number. Additionally, they found that higher Reynolds numbers and corrugation height along with lower corrugation helical pitch are better for HEX design, if heat transfer enhancement is targeted. The opposite is true if a lower pressure drop is desired. Nevertheless, their investigation was not extended to the transition region close to laminar flow regime. Recently, Darzi et al.~\cite{RABIENATAJDARZI2021106759} numerically investigated the thermal performance of coiled tubes having helical corrugated wall, while covering the transitional regime from laminar to turbulent flow. They developed a Nusselt number correlation (given as function of Dean number) based on the numerical results, which is found to be a stronger function of corrugation height compared to the helical pitch. 

\section{Experimental Investigation}
\label{sec:experimental}

In a previous research~\cite{ANDRADE2019940}, we provide a detailed literature survey on the experimental investigations on the heat transfer and pressure drop in helical corrugated tubes. In this section, we present the dimensionless numbers, and the available literature correlations to characterize the heat transfer and pressure drop in helical corrugated tubes under transitional flow regime.

\subsection{Characteristics of the tested tubes}
\label{sec:Characteristics of the tested tubes}

In the present experimental investigation, the pressure drop and heat transfer characteristics were measured in helical corrugated tubes with different helical pitches $p$ and the same inner diameter $D_i$ and corrugation height $e$. Water was used as the working fluid. According to the schematic representation of the corrugated tubes, as shown in Figure~\ref{fig:corrugated dimensions}, the characteristic dimensions of the investigated corrugated tubes (named P6, P9 and P12) are provided in Table~\ref{tab: tube dimensions}, where a smooth tube with the same inner diameter was also considered as the reference case. For all cases, the tube wall thickness $t$ was equal to 0.5~mm.

\begin{table}[!htbp]
\centering
\caption{Characteristic dimensions of the investigated helical corrugated tubes according to schematic representation depicted in Figure~\ref{fig:corrugated dimensions}, where smooth tube is used as reference.}
\begin{tabular}{lcccccccc} 
\hline
Tube & $D_i$ [mm]   & $D_h$  [mm] &  $A_c$ [mm$^2$] & $P_w$ [mm] & $A_s/L$ [mm] & $p$ [mm]     & $e$   [mm] &  $\phi$      \\
\hline
Corrugated - P6  &      &         &       &        & 14.31 & 6  &       & 0.006  \\
Corrugated - P9  & 4.5  & 4.35    & 15.06 & 13.85  & 14.06 & 9  & 0.4   & 0.004  \\
Corrugated - P12 &      &         &       &        & 13.97 & 12 &       & 0.003 \\
\hline
  Smooth         & 4.5  & 4.5     & 15.90 & 14.14  & 14.14 & $\infty$  & 0 & 0 \\
\hline
\end{tabular}
\label{tab: tube dimensions}
\end{table}

\begin{figure}[!htb]
\centering
  \includegraphics[trim=10 0 60 10,clip,width=0.4\textwidth]{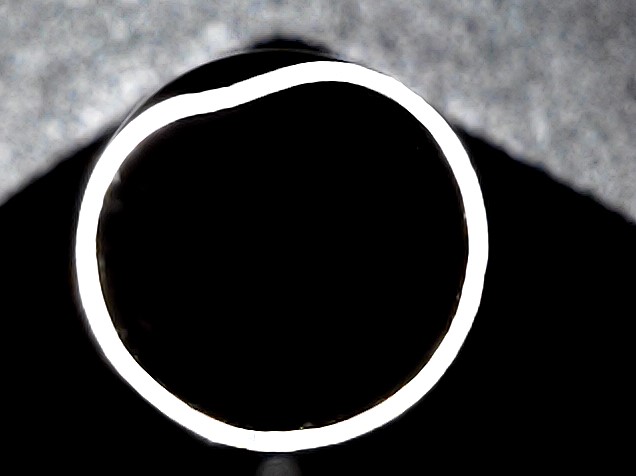}
  \caption{Microscope cross-section image of the corrugated tube P12.}
  \label{fig:micro cross section}
\end{figure}

For each corrugated tube, a microscopic cross-section image was obtained (see \textit{e.g.} the image of tube P12 in Figure~\ref{fig:micro cross section}), allowing the accurate estimation of the tube corrugation profile as well as the evaluation of the cross-section area $A_c$ and wetted perimeter $P_w$, which are required for calculating the hydraulic diameter $D_h$ through Eq.~\eqref{eq:hydro_diameter}. The obtained profiles were found to be nearly independent of the corrugation pitch and hence, the values of $A_c$ and $P_w$ were considered to be equal for all the three corrugated tubes considered in the present study (P6, P9 and P12); see Table~\ref{tab: tube dimensions}.

As it will be further detailed in section~\ref{sec:imple_comp_domain}, the obtained microscopic cross-section images were used to develop accurate CAD models for all investigated corrugated tubes; see for example the image of tube P12 in Figure~\ref{fig:corrugated real}. Based on this, the surface area of the inner tube wall per unit tube length $A_{s}/L$ was calculated. The values obtained through this procedure are also provided in Table~\ref{tab: tube dimensions}. It is worth mentioning that the wall surface area of corrugated tubes, where convective heat transfer takes place, increases as the helical pitch $p$ decreases and, for higher values of $p$ (\textit{i.e.} for P9 and P12), it can be smaller than the surface area of the smooth tube. This is not always the case but occurs for the present corrugated geometries since the value of $P_w$ is smaller than the one of smooth tube. Nevertheless, the relative variation of cross-section area and wetted perimeter, between all corrugated tubes and the reference case of smooth tube, never exceeds 2$\%$ and hence, it can be considered negligible for the present analysis of heat transfer enhancement.   

\subsection{Experimental setup}
\label{sec:experimental-setup}
\begin{figure}[!htb]
\centering
  \subfloat[Schematic setup overview]
    {\includegraphics[width=0.9\textwidth]{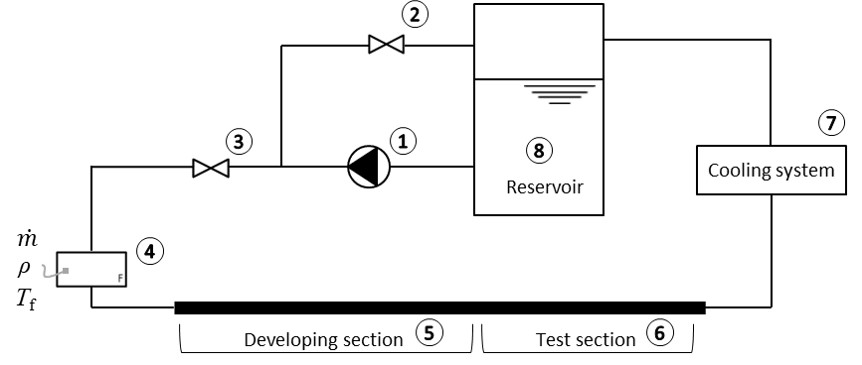}\label{fig:setup_overal}
    }\par
    \centering
    \subfloat[Detailed schematic view the of test section (6), showing the connection of pressure sensors and the location thermocouples: $x_{s,1} = 0.15$~mm; $x_{s,2} = 0.32$~mm; $x_{s,3} = 0.42$~mm; $x_{s,4} = 0.6$~mm]
    {\includegraphics[width=0.9\textwidth]{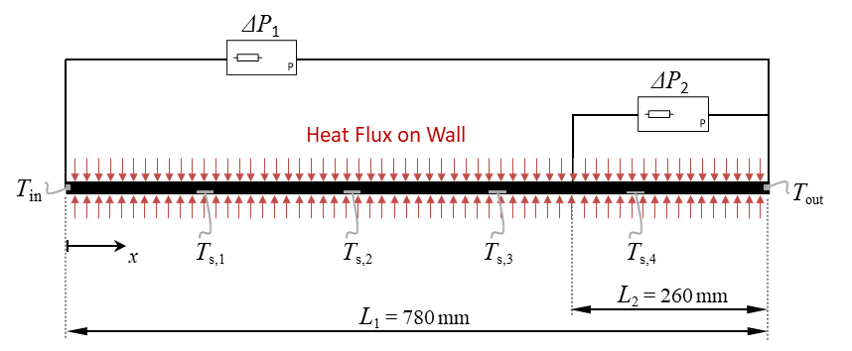}\label{fig:setup_test}
  }\par
  \caption{Schematic representation of the experimental setup, showing different components: (1) hydraulic pump; (2) and (3) flow regulation valves; (4) mass flow meter; (5) developing section; (6) test section; (7) cooling system; (8) reservoir.}
  \label{fig:setup1}
\end{figure}

The original experimental setup used in this work has been previously developed and used to investigate heat transfer enhancement in smooth tubes using nanofluids~\cite{Nikulin2019}. Subsequently, such setup has been further adapted and validated by Andrade et al.~\cite{ANDRADE2019940} to characterize pressure drop and heat transfer in corrugated tubes. The current version of the experimental setup is schematically represented in Figure~\ref{fig:setup1}. Compared to its previous version~\cite{ANDRADE2019940}, the length of the developing and test sections of the actual setup was increased in order to allow a proper comparison with numerical predictions in the fully developed flow regime, as it will be further discussed in section~\ref{sec:resul}. 

Regarding Figure~\ref{fig:setup1}(a), a magnetically coupled hydraulic vane pump (1) was used to force the working fluid to flow through the experimental setup. The pumping power was regulated by the pump speed, which, in turn was controlled by a frequency converter. Valve (2) was used to assist on the regulation of the desired flow, \textit{i.e.} when it is opened, it splits the flow rate, forcing the working fluid to recirculate back to the reservoir (8), thus allowing to adjust the flow rate. Valve (3) was also used for flow regulation purposes. The Coriolis mass flow meter (4) measured the mass flow rate $\Dot{m}$. Its working principle also allowed to obtain fluid density $\rho$ at known temperature values, also measured in the device. A platinum resistance thermocouple was installed on the mass flow meter, to measure the fluid temperature $T_{f}$; this thermocouple together with the fluid cooling system (7) play a paramount role in the setup, since they allow a constant control of the inlet fluid temperature. The developing section (5) consisted of a smooth tube with the length of 720~mm, to allow hydrodynamic fully developed flow profile under nearly adiabatic conditions. The test section (6), shown in detail in Figure~\ref{fig:setup1}(b) is 780~mm long and can easily accommodate any of the investigated tubes (smooth or corrugated) to be characterized. Downstream of the test section, the working fluid passes through the cooling system (7) and then returns to the fluid reservoir (8), closing the experimental loop.

A detailed schematic view of test section (6) is represented in Figure~\ref{fig:setup1}(b). The test section is heated by Joule effect, where a DC power supply (VOLTEQ HY5050EX ) provides a constant electric power (controlled by the imposed current) to the tube walls. This electric current is directly measured on the tube wall to account for the power dissipated through the conductive cables. In order to ensure minimal heat loses to surroundings, the tube of the test section and all associated connecting pipes were insulated with a porous rubber insulation (8~mm thick) with thermal conductivity lower than 0.04~W/mK. The pressure drop along the test section was measured using two pressure sensors connected to a differential pressure transducer. One ($ \Delta P_1$; a PX2300-100DI sensor), measured the pressure drop from the inlet to the outlet of the test section (through a tube length $L_1=780$~mm), while the other ($ \Delta P_2$; a PX2300-10DI sensor) measured the pressure drop through the last third of the test section length ($L_2=260$~mm). Using both sensors is a way of having a confirmation of the measurements using different sensitivity sensors. Along the test section, four type K thermocouples ($T_{s,1}$ to $T_{s,4}$) were assembled to the outer surface of the tube with a nearly even distance between them. Therefore, the local surface temperature was measured at four different locations in order to access the surface temperature evolution along the tube and hence, to clarify where the fully developed flow condition is reached within the test section. Moreover, two additional type K thermocouples were also installed to evaluate the average fluid temperature: one at the inlet ($T_{in}$) and the other at the outlet ($T_{out}$). A mixing chamber was added at the test section outlet, where the mixing of the fluid improves the accuracy of the measurement of $T_{out}$. Additional details on the experimental setup can be found in Andrade et al.~\cite{ANDRADE2019940}.

\subsection{Experimental data processing and uncertainty analysis}
\label{sec:experimental-data}
For the processing of the data measured at the test section, some thermo-physical properties of the working fluid (water) were required to be evaluated, namely \textit{i.e.} $c_p$, $\mu$ and $k$. These properties were estimated for water at its average temperature $\overline{T}$, defined as the mean value between the average fluid temperatures measured at the inlet and outlet of the test section. It is worth noting that while the measured average fluid temperature at the outlet was considered to be $T_{out}$, the average inlet fluid temperature $\overline{T}_{in}$ was evaluated as the mean value between $T_{in}$ and the fluid temperature $T_{f}$ measured in the mass flow meter.

Considering the fluid properties evaluated in the test section, both Reynolds and Prandtl numbers were calculated with Eqs.~\eqref{eq:reynolds} and~\eqref{eq:prandtl}. For the Reynolds number, the average velocity $u_m$ of the flow was obtained from Eq.~\eqref{eq:u_medio}, using the mass flow rate $\Dot{m}$ and fluid density $\rho$ measured in the mass flow meter as well as the cross-section area $A_c$ of the tube evaluated from the microscopic cross-section image.

The pressure drop measurements ($\Delta P_1$ and $\Delta P_2$) at different tube sections, allowed two different calculations of the Darcy friction factor $f$,using the following expression:
\begin{equation}
\ f=\frac{2}{\rho}\frac{\Delta P_i}{L_i}
\frac{u_{m}^2}{D_h} \:, \quad i = 1,2
    \label{eq:darcy}
    \end{equation}
where $\Delta P_i$ is the pressure drop measured along the respective length $L_i$, as shown in Figure~\ref{fig:setup1}(b).

Since a small part of the thermal power provided to the tube walls (through Joule effect) is dissipated to the surroundings, the heat rate $\dot{Q}$ effectively transferred to the fluid was calculated based on the following energy balance:
\begin{equation}
\dot{Q}=\dot{m} \, c_{p} \, \Delta T
\label{eq:q_fluid}
\end{equation}
where $\Delta T = T_{out} - \overline{T}_{in}$ is the fluid temperature variation between the inlet and outlet of the test section.

Note that the temperatures $T_{s,1}$ to $T_{s,4}$ were measured at the external surface of the tube. Therefore, the corresponding surface temperatures at the inner side of the tube ($T_{w,1}$ to $T_{w,4}$) were estimated accounting for the thermal resistance due to conduction through the tube wall:
\begin{equation}
T_{w,i} = T_{s,i}-\frac{\dot{Q}}{2 \pi \, L_1 \, k_{w}} \ln \left(\frac{D_{i}}{D_{i} + 2t}\right) \:, \quad i = 1,2,3,4
\end{equation}
where $k_w$ is the thermal conductivity of the tube wall material (stainless steel AISI 304), which was evaluated at the average surface temperature~\cite{InternationalHandbookCommittee1997}.

Considering that the heat rate $\dot{Q}$ is transferred to the fluid by convection, the local convective heat transfer coefficient $h$ can be calculated at different axial positions based on Newton's law:
\begin{equation}
h (x_i)=\frac{\dot{Q}}{A_s \left( T_{w,i}-\overline{T} \right)} \:, \quad i = 1,2,3,4
\label{eq:h exp}
\end{equation}
where $x_i$ is the axial position which corresponds to the temperature $T_{w,i}$, according to Figure~\ref{fig:setup1}(b) and $A_s$ is the tube surface area.

Using the trapezoidal rule approximation based on the set of values for local $h$, the average heat transfer coefficient $\overline{h}$ in the overall test section was determined from its definition: 
\begin{equation}
\overline{{h}}=\frac{1}{L} \int_{0}^{L} {h}(x) \cdot d x \approx \sum_{i=1}^{3} \frac{h\left(x_{i}\right)+h\left(x_{i+1}\right)}{2} \frac{\left[x_{i+1}- x_{i} \right]}{\left[x_{4}- x_{1} \right]}
\label{eq:aver h exp}
\end{equation}
Knowing the value of $\overline{{h}}$, it is straightforward to calculate the average Nusselt number using Eq.~\eqref{eq:Nusselt}.

Table~\ref{tab: ranges} summarizes the experimental range of the relevant variables, which are directly measured or post-processed based on the methodology described before. It is important to highlight that, for the different tube geometries, the experimental range of the Richardson number ${\rm Ri}$ = ${\rm Gr}/{\rm Re}^2$ (based on $D_h$ and $\Delta T$) is also included in the table. The magnitude of the Richardson number dictates the importance of natural convection with respect to forced convection and, since ${\rm Ri} < 0.1$ for all investigated cases, one can conclude that the dominant heat transfer mechanism in the present experiments is forced convection.

\begin{table}[!htb]
\centering
\caption{Experimental range of the relevant variables for the different investigated tubes.}
\begin{tabular}{lcccc}
\hline
Parameter & Smooth & Corrugated - P12  & Corrugated - P9      & Corrugated - P6  \\
\hline
$\overline{T}_{in}$~(ºC) & 24.5 - 26.3 & 24.0 - 26.3 & 24.1 - 25.9 &24.0 - 26.2 \\
$T_{out}$~(ºC) &29.5 - 35.9 & 29.1 - 35.3 &32.0 - 40.4 & 29.8 - 47.5 \\
$\Delta T$~(ºC) & 3.6 - 9.1 & 2.8 - 17.9 & 6.2 - 20.7  & 6.1 - 16.8  \\
$T_s$~(ºC) & 31.5 - 51.8 & 28.5 -46.3 & 29.8 - 45.8  & 30.0 - 51.1  \\
$\dot{Q}$~(W) & 45 - 247 & 57 - 355 & 57 - 364 & 59 - 291  \\
Pr & 5.35 - 5.78 & 4.84 - 5.83 & 4.84 - 5.45  & 5.00 - 5.59  \\
Re & 332 - 5646 & 315 - 4712 & 215 - 4929      & 303 - 4164  \\
$f$ & 0.0315 - 0.2379 & 0.1016 - 0.2135 & 0.1319 - 0.3098     & 0.1425 - 0.7172 \\
Nu & 5.5 - 41.9 & 3.8 - 76.1 & 2.8 - 63.1      & 3.5 - 81.0  \\
Ri & 0.000052 - 0.018 & 0.000038 - 0.018 & 0.000077 - 0.011 & 0.000081 - 0.021 \\
\hline
\end{tabular}
\label{tab: ranges}
\end{table}

The uncertainty analysis of the measurements was performed following the guidelines of Taylor and Kuyatt~\cite{Taylor1994}, where both random and systematic uncertainty errors are considered. The combined standard uncertainty of the post-processed experimental data was quantified following the procedure reported in previous studies, e.g. ~\cite{ANDRADE2019940,Nikulin2019}. Therefore, the reader is referred to these studied for further details. Table~\ref{tab: uncertainty} summarizes the standard uncertainties of the directly measured variables, which are required for evaluating the combined standard uncertainties of the post-processed variables; see Table~\ref{tab: uncertainty_output}.

\begin{table}[!htbp]
\centering
\caption{Standard uncertainty of directly measured variables (provided by the instrumentation specifications).}
\begin{tabular}{lcccc} 
\hline
Parameter & & & Uncertainty & Unit \\
 \hline
    Tube diameter, $D_i$  & &  & $\pm$ 0.05 & mm                          \\
    Tube length, $L$   & & & $\pm$ 3 & mm                    \\
    Inlet temperature, $\overline{T}_{in}$ & &  &  $\pm$ 0.2 & K                     \\
    Outlet temperature, $T_{out}$   & & &  $\pm$ 0.1 & K                     \\
    Surface temperature, $T_s$ & &  &  $\pm$ 0.1 & K                     \\
    Fluid density, $\rho$  & &  &  $\pm$ 5 & kg/m$^3$ \\
    Mass flow, $\dot{m}$ & & & $\pm$ 0.2 & \% \\
    Pressure difference, $\Delta P$ & & & $\pm$ 0.25 & \% \\
    \hline
\end{tabular}
\label{tab: uncertainty}
\end{table}

\begin{table}[!htb]
\centering
\caption{Maximum relative uncertainty for post-processed experimental variables}
\begin{tabular}{lcccc} 
\hline
Parameter & Smooth & Corrugated - P12  & Corrugated - P9      & Corrugated - P6        \\ 
\hline
Re & 1.3~\% & 1.4~\% & 1.7~\%  & 1.3~\%      \\
$f$ & 12.3~\% & 5.3~\%  & 10.0~\% & 6.5~\% \\
Nu & 20.0~\% & 24.2~\% & 19.6~\% & 22.6~\% \\
\hline
\end{tabular}
\label{tab: uncertainty_output}
\end{table}

\section{Numerical Modelling}
\label{sec:imple}

This section describes the 3D Computational Fluid Dynamics (CFD) model used for simulating the fluid flow and heat transfer inside the corrugated tubes considered in the present study (see Table~\ref{tab: tube dimensions}). This model was developed using the software Star-CCM+\textsuperscript{\textregistered}~\cite{star}.

\subsection{Computational domain for corrugated tubes}
\label{sec:imple_comp_domain}

The detailed geometry of real corrugated tubes is strongly dependent on the manufacturing process. Therefore, in order to minimize modeling inaccuracies, a direct generation of the computational domain in Star-CCM+\textsuperscript{\textregistered} software was avoided, based on schematic drawings of the tube geometry. Alternatively, 3D Computer-Aided Design (CAD) models were developed using the software Solidworks\textsuperscript{\textregistered}~\cite{solid} based on the real geometry of the experimentally investigated corrugated tubes, as illustrated by the microscopic cross-section shown in Figure~\ref{fig:micro cross section}. For each corrugated tube (P12, P9 and P6 in Table~\ref{tab: tube dimensions}), the obtained cross-section image was imported to the CAD software, allowing the accurate definition of the tube corrugation profile. The resulting profile was then swept along the tube length with a twist angle calculated from the corrugation helical pitch $p$. Additionally, to ensure that the computational domain integrally matches the real domain, the volume per unit length of the CAD model was verified against experimental values (which were obtained by measuring the mass of water that fills a tube of known length).

\begin{figure}[!htb]
\centering
  \subfloat[Cross-section view]
    {\includegraphics[width=0.4\linewidth]{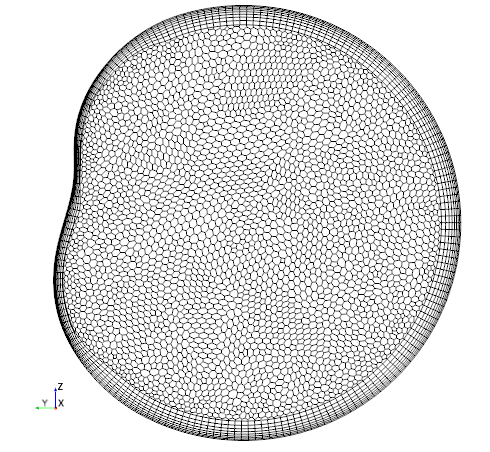}
    }\par
    \subfloat[Mid-section view]
    {\includegraphics[width=0.5\linewidth]{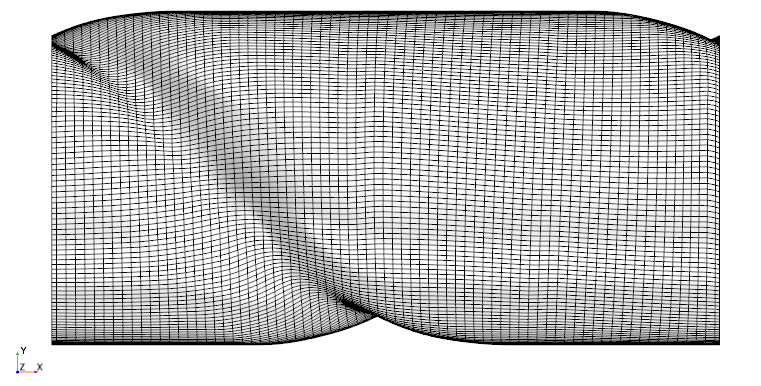}
  }\par
  \caption{Periodic computational domain of corrugated tube P12, including discretization mesh.}
  \label{fig:mesh example}
\end{figure}

The CAD geometry was imported to Star-CCM+\textsuperscript{\textregistered} software, considering the computational domain to be formed only by the fluid phase. For discretizing the computational domain, a polyhedral mesh was considered for the main fluid region, due to its high accuracy and low numerical diffusivity. Nevertheless, a set of 8 prismatic cell layers with an expansion ratio of 1.2 was applied close to wall boundaries in order to well-capture high spatial gradients of local velocity and temperature. The value of $y^+$ for locating the near wall grid nodes is below 1 for all simulated cases, thus excluding the need of wall functions for the turbulence modeling. As an example, the cross-section and mid-section views of a periodic computational domain are shown in Figures~\ref{fig:mesh example}(a) and~(b), respectively, where both polyhedral and prism layer meshes are visible. A grid independence study will be presented in section~\ref{section_verfi}.

\subsection{Governing equations and numerical method}
\label{CFD-Model}

The simulation of 3D incompressible flow and associated heat transfer inside the corrugated tubes requires mass, momentum and energy conservation equations to be numerically solved in the computational domain, while imposing proper boundary conditions. These equations can be expressed as:
\begin{eqnarray}
\label{continuity}
\nabla \cdot \textbf{v} &=& 0 \\
\label{momentum}
\rho \left( \frac{\partial \textbf{v} }{\partial t} + \textbf{v} \cdot {\nabla} \textbf{v} \right) &=& -{\nabla} P + \mu {\nabla}^2 \textbf{v} \\
\label{energy}
\rho \; c_p \left( \frac{\partial T }{\partial t} + \textbf{v} \cdot {\nabla} T \right) &=& k {\nabla}^2 T+\mu \Phi
\end{eqnarray}
where $\textbf{v}$ and $T$ are the local velocity vector and temperature, respectively, and $\Phi$ is the viscous dissipation function~\cite{bejan2013convection}.

For the present version of the governing equations, various simplifying assumptions were made, based on the experimental range of the variables shown in Table~\ref{tab: ranges}. The thermo-physical properties of the working fluid (\textit{i.e.} $\rho$, $\mu$, $k$ and $c_p$) were considered constant, which is a valid assumption since the maximum temperature variation of the fluid is lower than 20~K. Due to the low values of Richardson number (${\rm Ri}<0.1$), the heat transfer is dominated by forced convection and hence, buoyancy effects were neglected. Additionally, radiation heat transfer was also neglected since the system temperatures are close to the ambient one (\textit{i.e.} $< 50^\circ$C).

Using Star-CCM+\textsuperscript{\textregistered} software, the Finite Volume Method was applied to numerically solve the governing equations. Therefore, the transient term was discretized with a first-order approximation and the convection and diffusion terms were discretized by second-order upwind and central differencing schemes, respectively. A segregated solver was used together with a pressure-velocity coupling algorithm, \textit{i.e.} the Semi-Implicit Method Pressure Linked Equations (SIMPLE)~\cite{Patankar1980}.

Although the transient term in Eqs.~\eqref{momentum} and~\eqref{energy} was kept for the sake of generality, it was only considered for transient simulations in the laminar unsteady flow regime (close to the transition to turbulence). Alternatively, for both laminar steady and fully turbulent flow regimes, the steady-state version of the governing equations was rather solved using a steady solver in order to minimize computational costs. 

To close the governing equation system in the fully turbulent flow regime, the Reynolds-averaged Navier-Stokes (RANS) version of Eqs.~\eqref{continuity} to~\eqref{energy} was used~\cite{Patankar1980}. These alternative equations are similar to the governing equations for laminar regime but are rather expressed as function of mean flow transport quantities (\textit{e.g.} velocity and temperature), with the addition of extra terms that model the effects of flow fluctuations. For the present simulations, the transitional Shear Stress Transport (SST) model was chosen since, despite of its higher computational cost, it can deal with low Reynolds turbulence (close to the transition to laminar flow) and is suitable for modeling turbulent flow regions outside the boundary layer, such as swirl flows. This model has been previously applied~\cite{Haervig2017,XIN2018402} to simulate turbulent heat transfer in spirally corrugated tubes. Although recent works~\cite{CORCOLES2020106526,YANG2020106520} have also successfully applied the realizable $k-\varepsilon$ turbulence model in 3D simulations of heat transfer process in corrugated tubes, but at Reynolds numbers well above the transitional regime.

Due to the high computational costs of simulating the long tubes used in the experimental investigation, the entrance flow region was not considered for the present simulations and hence, only a periodic section of the corrugated tubes was modelled following a methodology typically applied in periodic flow problems~\cite{JORGE2019397}. In order to specify boundary conditions while preserving periodic flow conditions, Star-CCM+\textsuperscript{\textregistered} allows the creation of a periodic interface that maps a cycle with a rotation and translation. This interface is defined assuming hydrodynamically and thermally fully developed flow and it is created between two boundaries: the first, defined as a mass flow inlet in which the bulk inflow temperature can be specified, and a second, defined as a pressure outlet where the fully developed dimensionless velocity and temperature profiles are transposed to the inlet. Additionally, for the enclosing boundary of the computational domain defined by the tube walls, the no-slip boundary condition was imposed with an associated constant imposed heat flux. 

\subsection{Model verification and validation}
\label{section_verfi}

A critical judgement of the preliminary simulation predictions is required to access the confidence level of the results. Therefore, the model was verified and validated by "solving the equation right" and "solving the right equations", respectively ~\cite{Roache1997}.

\begin{figure}[!htb]
\centering
   \includegraphics[width=0.5\textwidth]{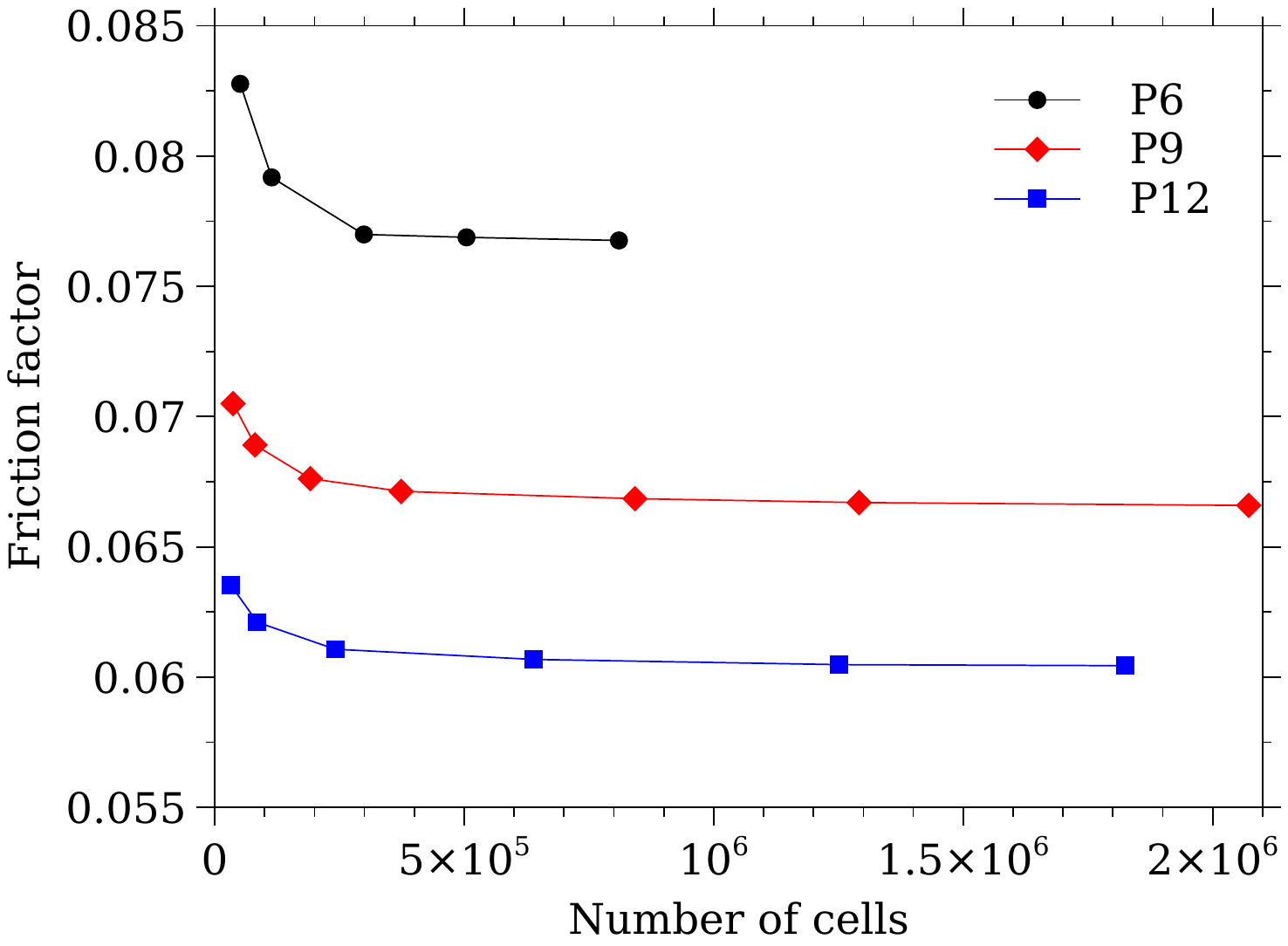}
  \caption{Grid independence study for the investigated corrugated tubes, showing the predicted friction factor at $Re=1300$ as function of the number of grid cells.}
  \label{fig:grid independe}
\end{figure}

Regarding model verification, a grid independence study was carried out for all investigated corrugated tube geometries (see Table~\ref{tab: tube dimensions}) and important quantities were calculated using computational meshes of different resolutions. Note that the prism layer thickness was kept constant during the grid refinement procedure. As an example, Figure~\ref{fig:grid independe} presents the grid independence study for a mass flow input that corresponds to ${\rm Re}=1300$ in the laminar flow regime. It can be concluded that, after a certain mesh resolution, there is no advantage in further increasing the number of grid cells, \textit{i.e.} the last refined mesh gives a relative difference of less than $0.5\%$ for the friction factor, when compared with the previous mesh. Based on this criterion, a mesh comprising around $1.2$ million grid cells was considered for corrugated tubes P12 and P9, whereas a mesh with $0.3$ million grid cells was used for tube P6.

For the model validation, a benchmark case of fully developed flow inside a periodic section of smooth tube was considered. Model predictions were found to be in excellent agreement with the results reported in the literature~\cite{bejan2003heat}. Furthermore, as it will be shortly discussed in section~\ref{sec:resul:valid}, the CFD model predictions for corrugated tubes were also compared against the experimental data obtained in this study.

\section{Results and discussion}
\label{sec:resul}
\subsection{Preliminary validation studies}
\label{sec:resul:valid}

The present experimental setup was firstly validated with the smooth tube, considering classical correlations reported in the literature. Based on the dimensionless Darcy–Weisbach friction factor $f$, the following correlations were considered for accessing the pressure drop for hydrodynamically fully developed flow: the Hagen-Poiseuille equation ($f = 64/Re$) and the Blasius equation ($f = 0.316 Re^{-1/4}$) for the laminar and turbulent regimes, respectively~\cite{bejan2003heat}, as well as the Hrycak and Andrushkiw correlation~\cite{Hrycak1974} for the transition region. For accessing the average convective heat transfer coefficient $\overline{{h}}$ inside the smooth tube, the Petukhov correlation~\cite{Petukhov1970} was considered for the entry flow region in laminar regime and, the Gnielinski correlation~\cite{bejan2003heat, Mohammed2013} was used for the turbulent regime assuming thermally fully developed flow.

\begin{figure}[!htb]
\centering
   \includegraphics[width=0.5\textwidth]{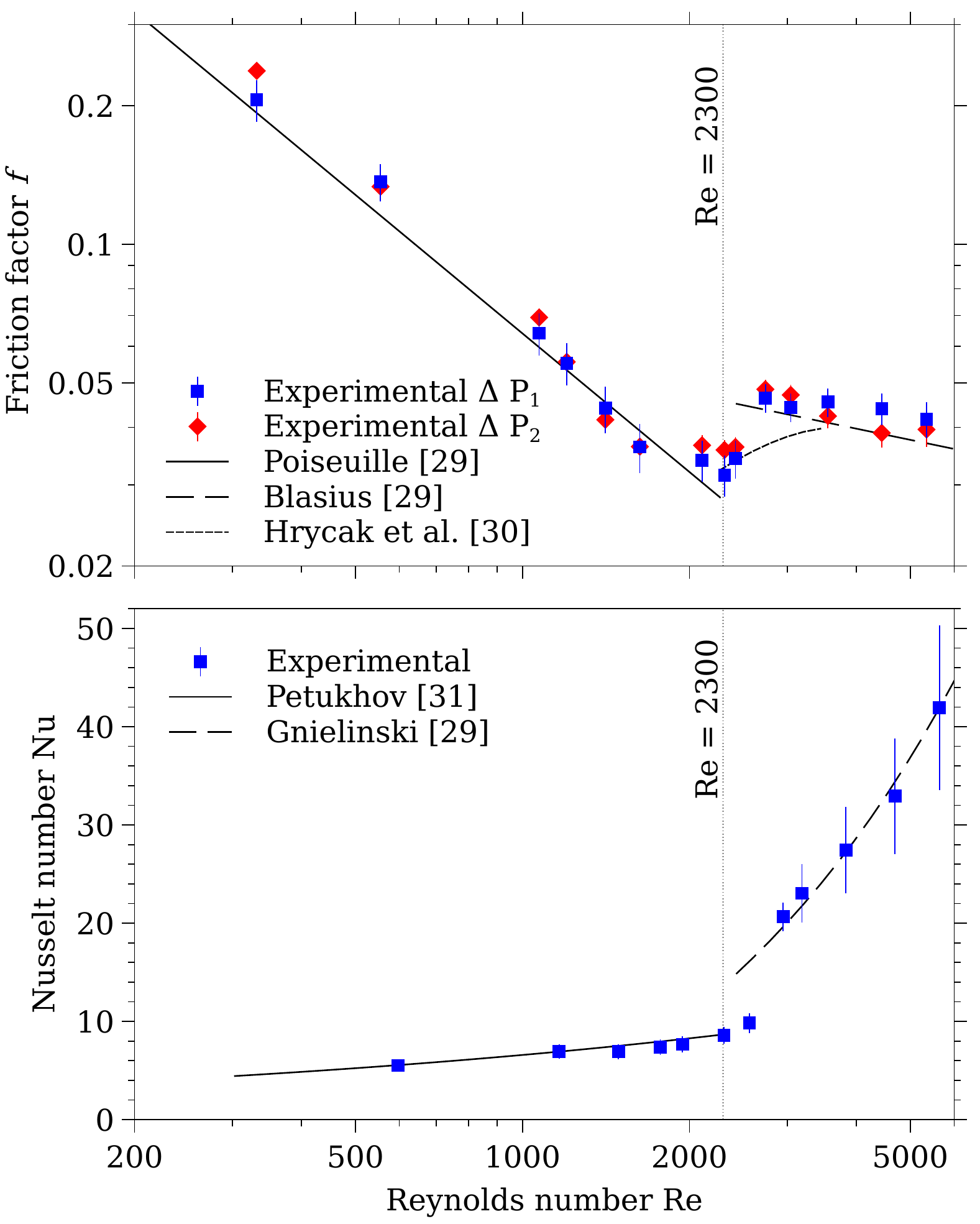}
  \caption{Validation of experimental setup with smooth tube, based on correlations from literature.}
  \label{fig:smooth_val_all}
\end{figure}

Figure~\ref{fig:smooth_val_all} shows that the friction factor and average Nusselt number measurements for smooth tube are in excellent agreement with the aforementioned correlations. For the Nusselt number comparison, the average relative error is less than 10\% for all the investigated range of Reynolds numbers. Similarly, the average relative error for the friction factor is under 12.5~\% for both laminar and turbulent regimes, although larger maximum relative errors (of the order of 20~\%) were found for the lowest Reynolds numbers tested in the laminar regime. Nevertheless, this discrepancy can be explained by small disturbances in the flow, which are known to strongly increase the friction factor for these laminar flow conditions, \textit{e.g.} welding defects on connections for pressure sensors and tube elements. In this respect, the errors on the friction factor measurements based on $\Delta P_2$ (see Figure~\ref{fig:setup_test}) tend to be higher than the ones for $\Delta P_1$, because small disturbances tend to be more amplified for shorter tube lengths and, this issue will be further highlighted in the discussion of the corrugated tube measurements provided below.

\begin{figure}[!htb]
\centering
   \subfloat[][Comparison of friction factor]
   {\includegraphics[width=0.5\linewidth]{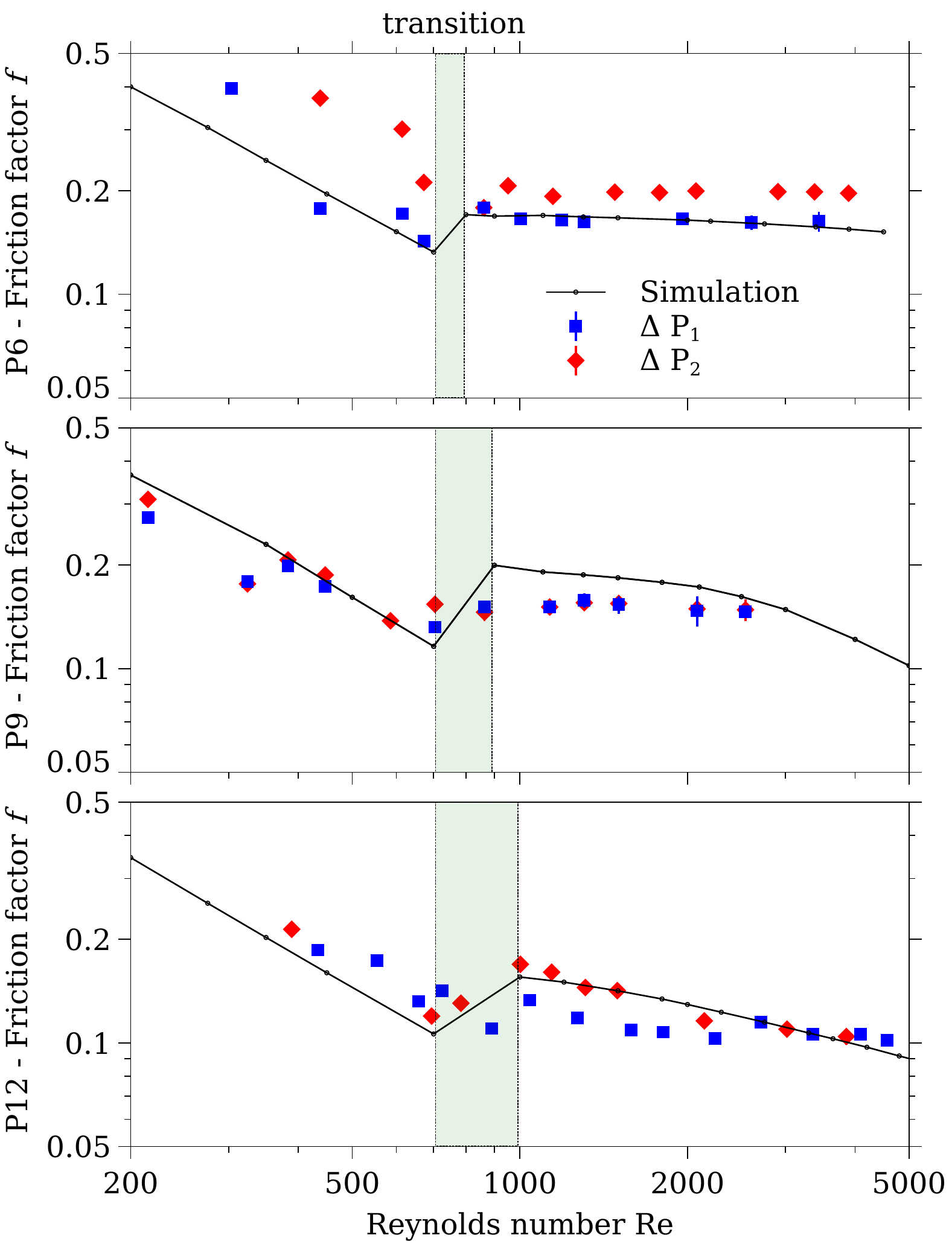}\label{fig:corr_val_f}}
   \subfloat[][Comparison of Nusselt number]
   {\includegraphics[width=0.5\linewidth]{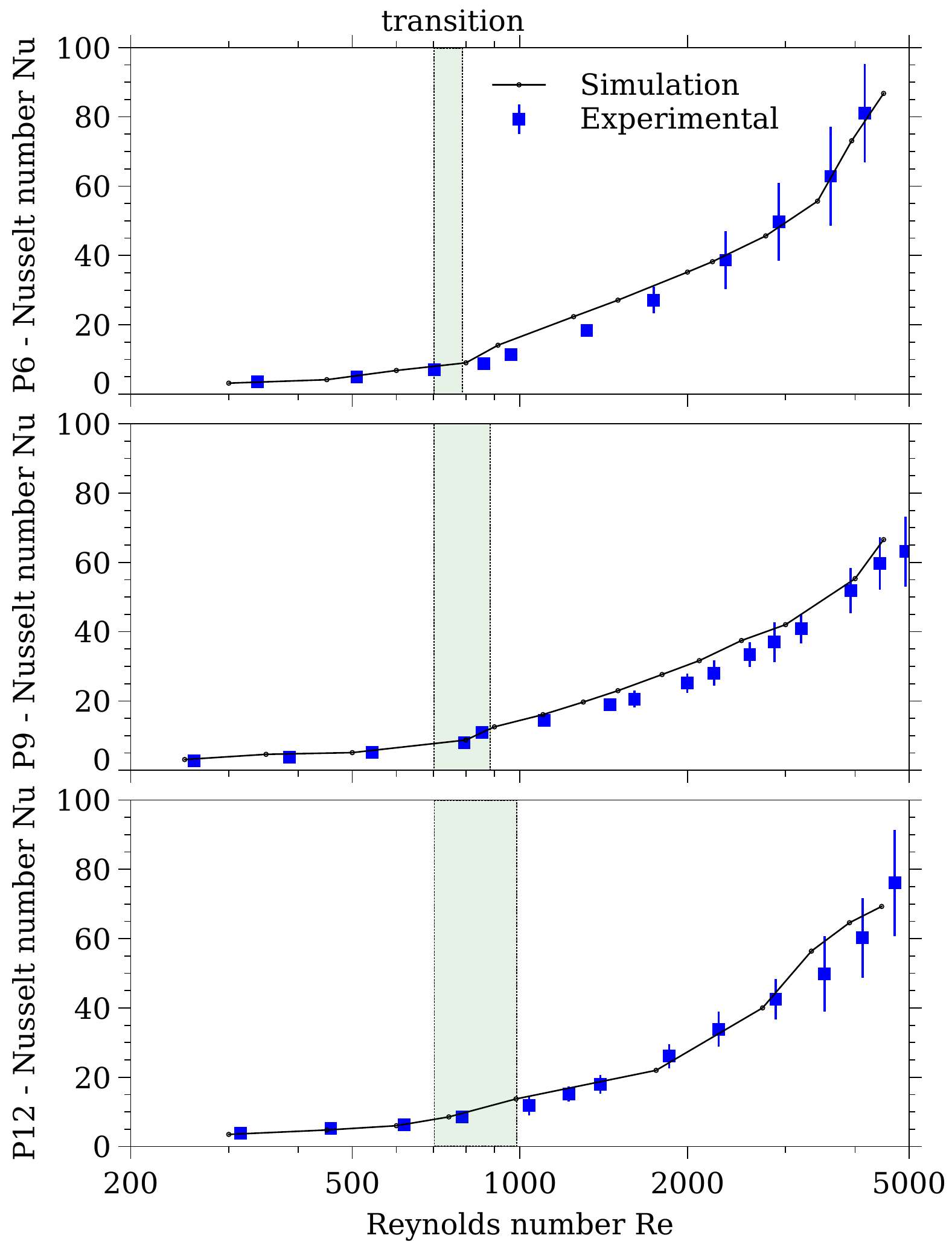}\label{fig:corr_val_Nu}}
  \caption{Validation of CFD model predictions with experimental results for all investigated corrugated tubes, covering a range of Reynolds numbers from laminar to turbulent regime.}
  \label{fig:corr_val}
\end{figure}

With the experimental setup validated, measurements for the corrugated geometries (P6, P9 and P12) were obtained and further used to validate and complement the CFD model predictions. Regarding the friction factor and Nusselt number for all investigated corrugated tubes, a comparison between the experimental data and model predictions is presented in Figures~\ref{fig:corr_val_f} and~\ref{fig:corr_val_Nu}, respectively. It is important to clarify that, since the model predictions are obtained for fully developed flow, the Nusselt number measurements used for comparison are the ones acquired in the last part of the test section (see $T_{s,4}$ in Figure~\ref{fig:setup_test}). In this respect, the measurements for laminar flow regime were obtained under thermally developing flow conditions for a significant initial part of the test section before approaching fully developed conditions. On contrary, around 86~\% of the test section length was found to be thermally fully developed in the turbulent flow regime. Regarding the experimental friction factors presented in Figure~\ref{fig:corr_val_f}, one can observe that both $\Delta P_1$ and $\Delta P_2$ measurements provide similar values of $f$ for P9 and P12 corrugated tubes. However, it is clear that for P6 geometry the values of $f$ obtained from $\Delta P_2$ are overestimated, probably due to some welding defects on the intermediate pressure sensor connection. Therefore, this experimental data set should be disregarded for model validation purposes.  


Overall, Figure~\ref{fig:corr_val} shows a very good agreement between the experimental and numerical results, which validates the numerical model described in section~\ref{sec:imple}, particularly the assumptions considered for the internal flow. Taking into account that the CFD model is able to well-capture the pressure drop and heat transfer behaviors inside corrugated tubes for all the investigated range of flow regimes, the obtained predictions can provide important phenomenological details of the process, as further discussed in the following sections.

\subsection{Pressure drop behavior}
\label{sec:resul:pressure}

One can observe from Figure~\ref{fig:corr_val_f} that the corrugated geometries promote the laminar-turbulent transition to occur at lower Reynolds numbers than that for a smooth tube (where ${\rm Re}_{cr} \approx 2300$). These results are in agreement with those reported in previous studies~\cite{ANDRADE2019940,Vicente2004,Vicente2004a,MEYER20111587} and, they suggest that the transition is mainly governed by the corrugation height, starting at ${\rm Re}_{cr} \approx 700$ for all investigated corrugated tubes, independently of their corrugation pitch $p$ (see Table~\ref{tab: tube dimensions}). Having the same corrugation height associated to a similar value of ${\rm Re}_{cr}$ for P6, P9 and P12, this remark is of great importance for the characterization of the flow inside corrugated geometries, as further discussed in section~\ref{sec:resul:charact}.

As the Reynolds number increases in the turbulent regime, the model predictions from Figure~\ref{fig:corr_val_f} reveal that the friction factor for the corrugated tubes P9 and P12 clearly tends to decrease, which is in good agreement with previous results reported in the literature~\cite{Garcia2012,Meyer_2014}. On the other hand, the tube P6 has a friction factor nearly constant along the turbulent regime (${\rm Re} \gtrsim 800$). This limit behaviour can be explained by its high severity index ($\phi$ = 0.006), which precludes the development of the laminar sub-layer of the boundary layer in a similar way to what occur for rough pipe flow.

\begin{figure}[!htb]
\centering
   \includegraphics[width=0.5\linewidth]{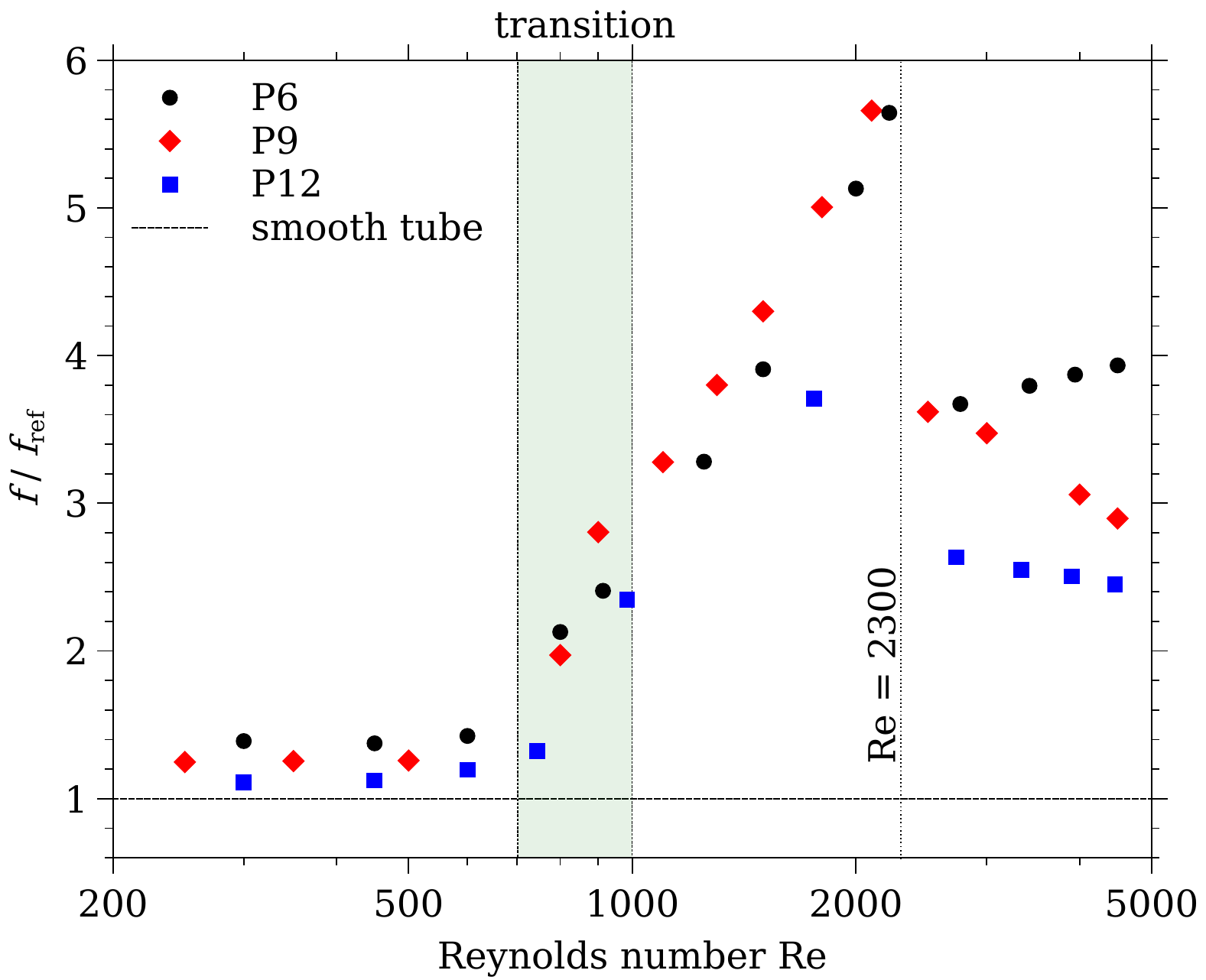}
  \caption{Comparison of model predictions for the friction factor $f$ of corrugated geometries against the reference value $f_{ref}$ of smooth tube obtained from classical correlations, covering a range of Reynolds numbers from laminar to turbulent regime.}
  \label{fig:f_compare}
\end{figure}

Considering that similar conclusions can be extracted from both experimental and numerical data in Figure~\ref{fig:corr_val_f}, the pressure drop performance of corrugated geometries was compared with the one of the smooth tube based only on the CFD model results. Figure~\ref{fig:f_compare} depicts the numerical predictions of friction factor ratio $f/f_{ref}$ as function of Reynolds number, where $f$ for corrugated tubes (P6, P9 and P12) is made dimensionless with respect to $f_{ref}$ obtained from classical correlations for the smooth tube, \textit{i.e.} Hagen-Poiseuille and Blasius correlations for laminar and turbulent regimes, respectively (see section~\ref{sec:resul:valid}).

One can observe from Figure~\ref{fig:f_compare} that, depending on the interest range of Reynolds numbers, the helical corrugation can cause a significant increase on the friction factor, when compared with the one of smooth tube. In this respect, all the corrugated tubes present the expected behavior: the laminar-turbulent transition occurs at lower Reynolds numbers ($700 \lesssim Re \lesssim 1000$) with an associated "jump" of the friction factor. Therefore, the helical corrugation can be regarded as a mean to promote the transition to the turbulent regime, while keeping a low mass flow rate of the working fluid. 

Figure~\ref{fig:f_compare} also shows that, in the laminar regime, the corrugated geometries yield a slight increase of the friction factor due to an increase of the wall friction drag, caused in part by the reduction of the cross-sectional area available for the flow. In this regard, the corrugated tube with the lowest helical pitch (P6) presents the highest friction factor value. For an intermediate range of Reynolds number in the turbulent regime ($900 \lesssim Re \lesssim 2300$), the corrugated tube P9 yields a slightly higher friction factor, which is then overpassed again by the corrugated tube P6 for higher Reynolds numbers. Overall, the helical corrugation acts as an artificial roughness that, as the Reynolds number increases, tends to be out of the boundary layer and hence, is more subjected to the main flow. Consequently, the friction factor induced by normal stress is intensified by reducing the pitch of corrugations. Therefore, the corrugated tube that presents the lowest friction factor is the P12, \textit{i.e.} the one with the highest pitch (see Table~\ref{tab: tube dimensions}).

\subsection{Flow characterization}
\label{sec:resul:charact}

\begin{figure}[!htb]
\centering
\subfloat[Laminar regime: Re = 500]
   {\includegraphics[width=0.5\linewidth]{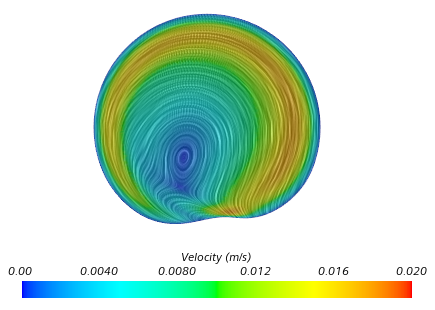}\label{fig:tang_vel} }
   \subfloat[Turbulent regime: Re = 2100]
   {\includegraphics[width=0.5\linewidth]{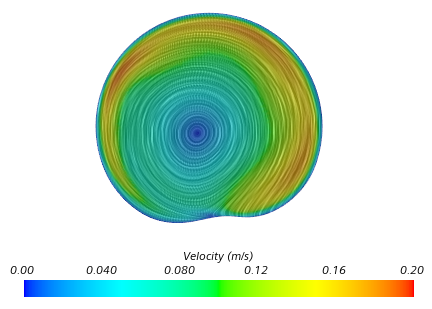}\label{fig:tang_vel_turb} }
   \par
  \caption{CFD model prediction of tangential velocity component and stream lines projection, showing cross-section scalar view for corrugated tube P9 at different flow regimes.}
  \label{fig:tang_vel_all}
\end{figure}

Based on the detailed analysis of the CFD model predictions for all corrugated geometries, one can complement the previous phenomenological interpretation of pressure drop behaviour and clarify its association to the promoted heat transfer enhancement. In this respect, it was observed that the effect of the tube corrugation on the tangential velocity component is more significant than on the radial component. Figure~\ref{fig:tang_vel_all} depicts the tangential velocity component for the corrugated tube P9 at laminar and turbulent flow regimes, providing a cross-section scalar view including the projection of stream lines. From Figure~\ref{fig:tang_vel} for laminar regime (Re~=~500), one can observe that the induced tangential velocity promotes the flow to swirl with more intensity close to the tube walls and towards the corrugation, where the stream lines show some local ineffectiveness of the swirl effect. This local ineffectiveness of the swirl tends to vanish for turbulent flow, as shown in Figure~\ref{fig:tang_vel_turb} for Re~=~2100. The relevance of the tangential velocity on the overall flow is noticed by the fact that its maximum magnitude ranges from around 10 to 30~\% of the axial velocity magnitude as the flow goes from laminar to turbulent, which adds to the well-distributed presence of this velocity component in a large fraction of the cross-section area. This result shows a key effect of the tube corrugation on the flow: the tangential velocity generated by the corrugation induces a swirl in the flow, which promotes a heat transfer enhancement.

Based on the swirl number $S_m$ defined by Eq.~\eqref{eq:swirl_number}, the CFD results were post-processed and the values of $S_m$ for the different corrugated tubes (P6, P9 and P12) under different flow regimes (laminar at Re~=~500 and turbulent at Re~=~2100) are provided in Table~\ref{tab:swirl_number}. One can conclude from the table that a higher value of swirl number is obtained for an intermediate corrugation pitch and that this value increases substantially as the flow becomes turbulent. This can be explain by the fact that the helical corrugation acts in two different ways: it promotes swirl as well as it creates artificial roughness where the corrugation obstacle induces boundary layer disturbances in the axial flow direction. For high pitches, the helical corrugation tends to be more aligned with the axial flow direction and hence, its effectiveness in creating swirl is reduced. On the other hand, for low pitches, the corrugation mainly acts as an obstacle, which tends to increase the recirculation zone that dissipates the swirl effect.

\begin{equation}
S_m = \frac{\int_{A_c} u_{tang} \; u_{axial} \; dA_c}{\int_{A_c} u_{axial}^2 \; dA_c}
\label{eq:swirl_number}
\end{equation}

\begin{table}[!htb]
\centering

\caption{CFD model prediction of swirl number $S_m$ for different corrugated tubes under different flow regimes.}
\begin{tabular}{lcccc} 
\hline
Tube & Re = 500 & Re = 2100  \\ 
\hline
P12 & 0.0342 & 0.105 \\
P9 & 0.0464 & 0.158 \\
P6 & 0.0417 & 0.124 \\
\hline
\end{tabular}
\label{tab:swirl_number}

\end{table}

\begin{figure}[!htb]
\centering
   \subfloat[Laminar regime: Re = 500]
   {\includegraphics[width=0.5\linewidth]{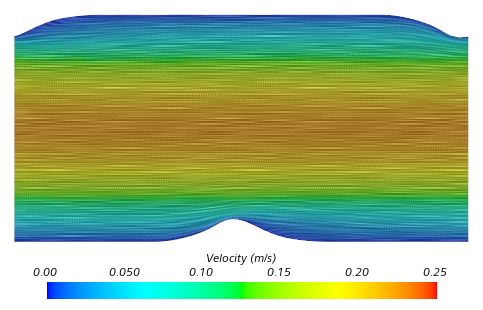}\label{fig:sep_lam} }
   \subfloat[Turbulent regime: Re = 2100]
   {\includegraphics[width=0.5\linewidth]{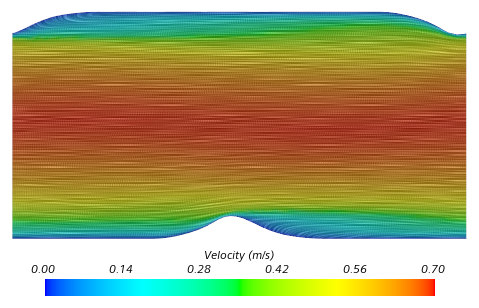}\label{fig:sep_turb} }
   \par
  \caption{Mid-section plane with velocity magnitude and stream lines projection for corrugated tube P9 at different flow regimes.}
\end{figure}

With the aim of providing a deeper analysis on the flow characterization and better sustain these arguments, Figure~\ref{fig:sep_lam} depicts the mid-section plane for P9 geometry at Re~=~500, where the behavior of the axial velocity and stream lines can be observed for the laminar regime slightly before transition. It is possible to conclude that the flow occurs with a negligible separation caused by the corrugation obstacle, where a low velocity region is stabilized and promotes smooth velocity gradients at the near wall. 

For comparison purposes, similar information is presented in Figure~\ref{fig:sep_turb} for P9 at ${\rm Re} = 2100$ (turbulent regime). By analysing Figure~\ref{fig:sep_turb} with respect to Figure~\ref{fig:sep_lam},  one can notice that in the fully turbulent regime there is a clear flow separation caused by the corrugation, which promotes a downstream recirculation and subsequent reattachment of the flow. Additionally, the higher velocities of the core tend to propagate to the near wall region promoting stronger velocity gradients. This analysis shows the relevance of the corrugation height in the flow and how this design dimension can be used to control the transition region and consequently, the operating regime for a desired mass flow rate.

In Table~\ref{tab:recirculation_zone}, the extension of the recirculation zone is quantified for the all simulated corrugated tubes and for two Reynolds numbers (Re~=~500 and Re~=~2100) representative of laminar and turbulent flow regimes. One can conclude from the table that lower corrugation pitches and higher Reynolds numbers promote a more pronounced recirculation zone, however, under opposite conditions there is a negligible or even absent recirculation zone (see e.g. Figure~\ref{fig:sep_lam}).

\begin{table}[!htb]
\centering

\caption{CFD model prediction of dimensionless recirculation zone length (taking the corrugation pitch as reference) for different corrugated tubes under different flow regimes; Note: N/A stands for absence of well-defined recirculation zone.}
\begin{tabular}{lcccc} 
\hline
Tube & Re = 500 & Re = 2100  \\ 
\hline
P12 & N/A & N/A \\
P9 & N/A & 0.28 \\
P6 & 0.17 & 0.50 \\
\hline
\end{tabular}
\label{tab:recirculation_zone}

\end{table}

\begin{figure}[!htb]
\centering
\subfloat[Local Nusselt number]
   {\includegraphics[trim = 185 0 270 0, clip, width=0.35\linewidth]{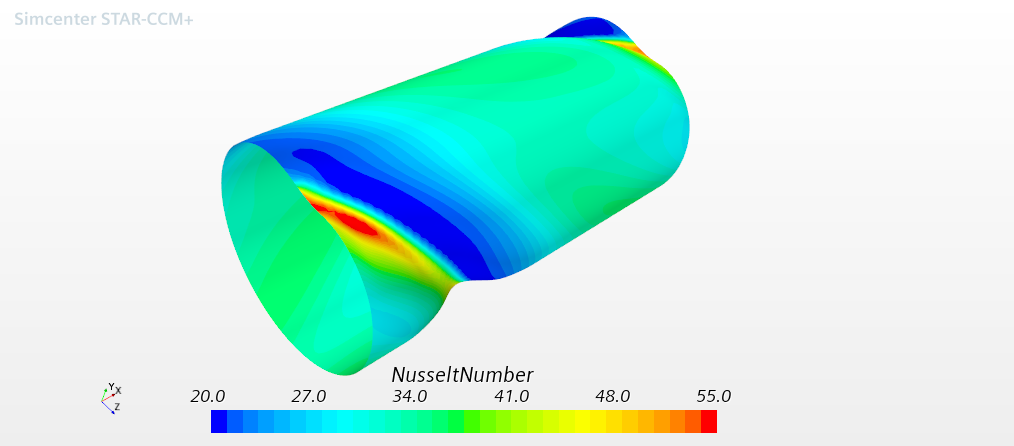}\label{fig:nu_turb_wall} }
   \subfloat[Local dimensionless temperature]
   {\includegraphics[trim = 185 0 270 0, clip, width=0.35\linewidth]{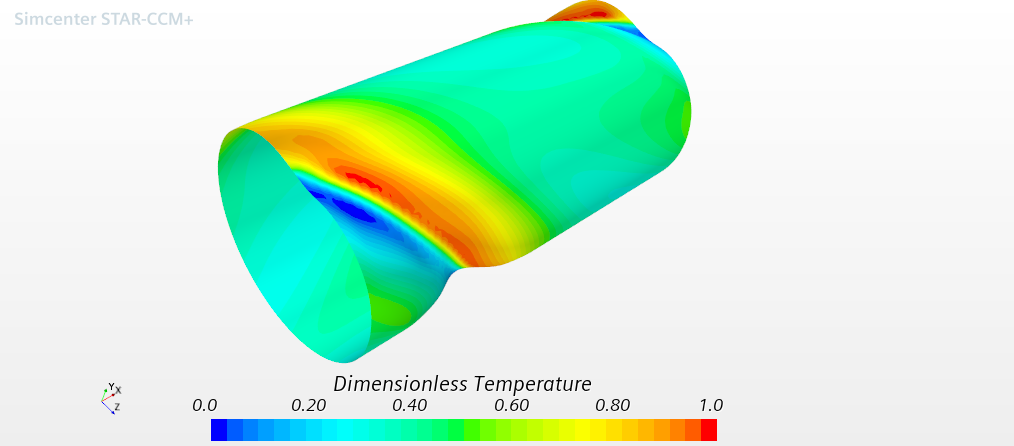}\label{fig:temp_turb_wall} }
   \par
  \caption{Distribution of local heat transfer characteristics at tube wall for P9 at $Re =2100$ (turbulent regime).}
  \label{fig:heat_turb_wall}
\end{figure}

In association to Figure~\ref{fig:sep_turb}, the local Nusselt number distribution at the tube wall is shown in Figure~\ref{fig:nu_turb_wall} for the turbulent regime. One can observe that the highest Nu is found upstream of the corrugation peak. In the recirculation region, the value of Nu is found to be low due to the low velocity magnitude, however the free shear layer generated by the recirculation enhances the further evolution of the Nu number. Moreover, it is interesting to note that a large Nu variation is also observed at the corrugation obstacle and follows the corrugation path along the pitch. The Nusselt variation along the periodic domain of the flow provides the important understanding that the flow is thermally pulsating, which is in agreement with the experimental observations from Andrade et al.~\cite{ANDRADE2019940}. Additionally, this spatial variation of Nusselt number is important in the design of corrugated tubes, as it allows one to identify wall regions with large temperature gradients and possible associated hot spots. In this respect, Figure~\ref{fig:temp_turb_wall} depicts the dimensionless temperature, scaled between the respective minimum and maximum values at the wall of the periodic domain.

\subsection{Heat transfer enhancement}
\label{sec:resul:heat}

This section aims to compare the increase in heat transfer coefficient produced by the corrugated tubes in the laminar, transitional and turbulent flow regimes. To this end, experimental and numerical results can provide similar conclusions, as demonstrated in Figure~\ref{fig:corr_val_Nu}. Hence, only CFD model predictions will considered for the following discussion.

\begin{figure}[!htb]
\centering
   \includegraphics[width=0.5\linewidth]{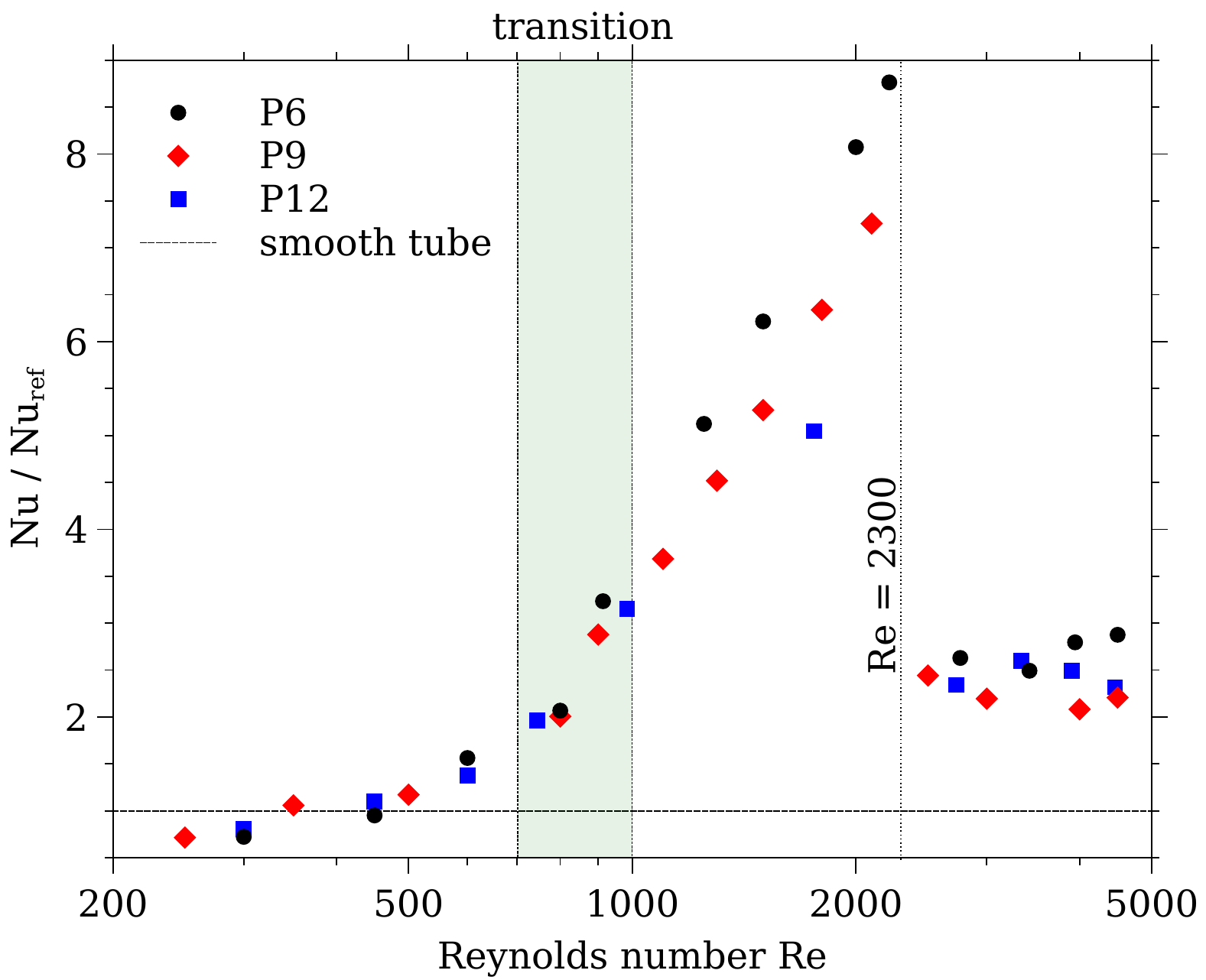}
  \caption{Nusselt number predictions for the corrugated tubes compared against the reference value ${\rm Nu}_{ref}$ for smooth tube obtained from classical correlations, covering a range of Reynolds numbers from laminar to turbulent regime under fully developed flow conditions.}
  \label{fig:nu_compare}
\end{figure}

Figure~\ref{fig:nu_compare} shows the Nusselt number ratio ${\rm Nu}/{\rm Nu}_{ref}$ between corrugated and smooth tube for all investigated geometries (P6, P9 and P12) as function of Reynolds number, where ${\rm Nu}_{ref}$ for the smooth tube is obtained from classical correlations under imposed wall heat flux and fully developed flow conditions: ${\rm Nu}_{ref} = 4.36$ for laminar regime and, the Gnielinski correlation for turbulent regime~\cite{bejan2003heat}.

One can observe from Figure~\ref{fig:nu_compare} that in the laminar regime (${\rm Re} \lesssim 700$) the Nusselt number for all corrugated tubes is comparable (slightly lower or higher) to that obtained for the smooth tube, which indicates that the flow field in the corrugated tubes is only affected to a minor degree. This can be explained by the small swirl effect induced by the corrugation, while the flow stays attached to the wall; see Figure~\ref{fig:sep_lam}. Therefore, it can be concluded that in terms of heat transfer enhancement, the application of corrugated tubes in the laminar flow regime is not very effective.

On the other hand, Figure~\ref{fig:nu_compare} also reveal that, as the Reynolds number increases towards the transition to turbulent flow ($700 \lesssim {\rm Re} \lesssim 1000$), the corrugated tubes promote a significant increase of the Nusselt number, when compared to the constant ${\rm Nu}$ value for smooth tube under laminar flow (\textit{i.e.} for ${\rm Re} < 2300$). This can be explained by the fact that, besides promoting an early laminar-turbulent transition, the helical corrugation induces a more severe swirl effect as the Reynolds number increases. It is important to highlight that, for Reynolds numbers up to the end of laminar-turbulent transition, all investigated corrugated geometries yield very similar ${\rm Nu}$ values.  

After turbulent flow has been established for corrugated geometries (${\rm Re} \gtrsim 1000$), Figure~\ref{fig:nu_compare} shows that the highest Nusselt number occurs for geometry P6 that presents the lowest pitch, \textit{i.e.} the highest severity index. As discussed in section~\ref{sec:resul:charact}, this is justified by the dominant effect of the corrugation obstacle on the boundary layer at high Reynolds numbers. More specifically, the mechanism of Nusselt number enhancement is based on a multi-mode flow phenomena, where the swirl effect is superposed to the flow separation and reattachment induced by the corrugation obstacle; see Figure~\ref{fig:nu_turb_wall}. Due to the flow reattachment, the Nusselt number enhancement rapidly deteriorates as the boundary layer grows. Nevertheless, the corrugation obstacle promotes periodic redevelopment of this boundary layer, which is intensified by a lower corrugation pitch (\textit{i.e.} higher roughness), therefore causing a more effective heat transfer~\cite{Mohammed2013}. 

Overall, a significant heat transfer enhancement (up to $\sim 8$ times that of smooth tube) can be achieved with the investigated corrugated tubes. However, one can conclude from Figure~\ref{fig:nu_compare} that, for ${\rm Re} \gtrsim 2300$, the Nusselt number enhancement caused by the corrugated tubes falls down and tends to stabilize at a value of 2-3 times that of the smooth tube. As expected, this demonstrates that the benefit of employing corrugated tubes for increasing heat transfer is less significant after the flow inside the smooth tube becomes turbulent.

\section{Thermal performance factor}

Based on the previous separate discussion for the friction factor $f$ and Nusselt number ${\rm Nu}$, a further relevant analysis addresses the comparison between all investigated corrugated tubes and the reference case of smooth tube through the thermal performance factor $\eta$, defined according to Eq.~\eqref{eq:thermal performance}. This parameter accesses the combined positive/negative effects of the corrugated geometries with respect to both heat transfer and pressure drop, while considering a pumping power equal to the one of smooth tube. Since the reference case of smooth tube is assumed to suddenly transit from laminar to turbulent at ${\rm Re} = 2300$, any quantitative ratio between corrugated and smooth tubes (including $f$, ${\rm Nu}$ or $\eta$) is expected to be affected by reflecting a similar jump.



\begin{figure}[!htb]
\centering
   \includegraphics[width=0.5\linewidth]{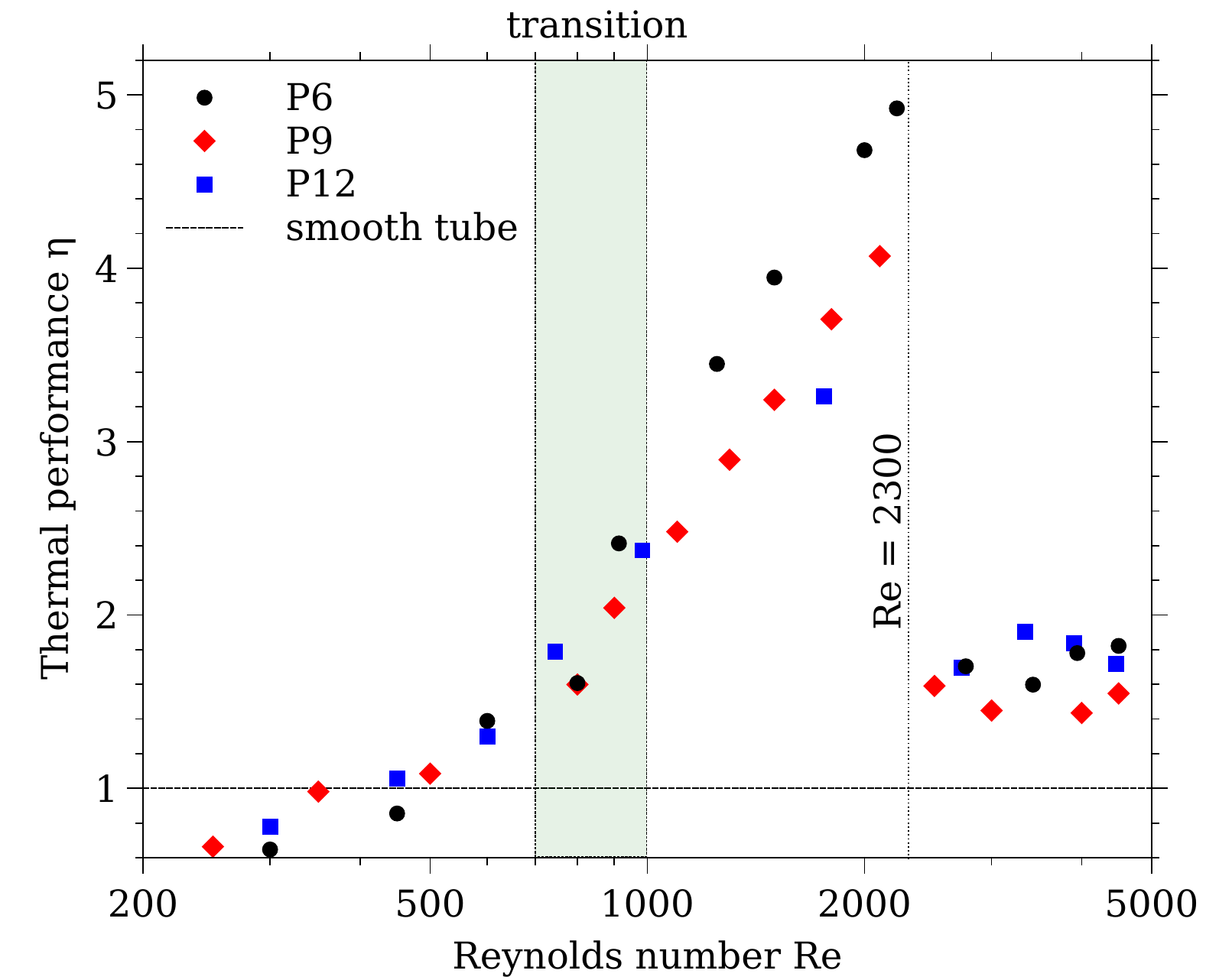}
  \caption{Thermal performance factor of all investigated corrugated tubes, based on the CFD model results.}
  \label{fig:thermal perfor}
\end{figure}

Figure~\ref{fig:thermal perfor} presents the numerical results for the thermal performance factor as function of Reynolds number, for all corrugated tubes. For laminar flow (${\rm Re} \lesssim 700$), one can observe that the behavior of $\eta$ is similar to the one of ${\rm Nu}$ since $f$ is nearly constant (see Figures~\ref{fig:f_compare} and~\ref{fig:nu_compare}). In this respect, all corrugated geometries yield similar thermal performance factors that increase with the Reynolds number up to a value of $\sim 1.5$. 

One can note from Figure~\ref{fig:thermal perfor} that the aforementioned increase of $\eta$ continues as long as the laminar regime remains for the smooth tube (${\rm Re} \lesssim 2300$). However, after the flow in the corrugated tubes becomes turbulent (${\rm Re} \gtrsim 1000$), the geometry with the lowest pitch (P6) tends to yield a clearly higher thermal performance factors (maximum $\eta$ of around $5$ for ${\rm Re} \approx 2300$). In part, this is associated to the similar effect observed for ${\rm Nu}$ (see Figure~\ref{fig:nu_compare}) but also to the negative effect of $f$ for P9 (Figure~\ref{fig:f_compare}), making $\eta$ values for P9 and P12 geometries to compete with each other. After having the discontinuity with a sudden decrease of $\eta$ at ${\rm Re} \approx 2300$ (\textit{i.e.} the flow in smooth tube becomes turbulent), the effect of $f$ for P9 becomes less negative but the opposite occurs to its ${\rm Nu}$ value (see Figures~\ref{fig:f_compare} and~\ref{fig:nu_compare}). Therefore, although $\eta$ remain in the range of $1.5-2$ for all corrugated tubes, the geometry P9 with the intermediate pitch tends to yield a lower $\eta$ for this range of Reynolds numbers in the turbulent regime.

Overall, one can conclude that for a fixed corrugation height, the corrugated tube with lowest pitch (P6) performs better in terms of thermal performance for practical applications, where the optimal Reynolds number range of operation lays between 1000 and 2300. Nevertheless, it is worth to clarify that the better performance of P6 is associated to its higher severity index, whose parametric variation was solely imposed by the range of corrugation pitches considered in this investigation. Therefore, based on the validated CFD model, further parametric studies could be perform in future works in order to better access the optimal geometrical parameters and range of operating conditions that maximize the thermal performance of helical corrugated tubes, including for example the corrugation shape.  

\section{Conclusions}
\label{sec:concl}

The present work addresses the characterization of fluid flow and heat transfer inside corrugated tubes for an imposed wall heat flux, while covering a range of Reynolds numbers from the laminar to turbulent regime. Three helical corrugated geometries, with different corrugation pitches, were modelled and simulated using a commercial CFD software for predicting the three-dimensional flow velocity and temperature fields, as well as the overall pressure drop and heat transfer behaviors. An experimental setup adapted from a previous work~\cite{Nikulin2019} was used to test those corrugated geometries and to validate the CFD model predictions. Additionally, a smooth tube was used to validate the experimental data against classical correlations reported in literature and, it was also taken as reference for accessing the performance of the tested corrugated geometries with respect to a quantitative thermal performance factor. 

The main conclusions extracted from this work can be summarized as follows:
\begin{itemize}
    \item For all the investigated range of Reynolds numbers, the CFD model predictions are found to be in good agreement with the experimental results for both friction factor and Nusselt number.
    
    \item The friction factor measurements clearly show that the transition region is anticipated for the corrugated tubes, when compared with the smooth tube. Nevertheless, the critical Reynolds number for the start of transition is found to be approximately the same for the various corrugated geometries. Additionally, based on the simulation results, a further analysis of the flow characteristics suggests that the transition is governed solely by the corrugation height, which is in agreement with the previous work from Vincent et al.~\cite{Vicente2004}.
    
    \item From the velocity vector field predicted for the corrugated geometries, it can be demonstrated that the helical corrugation shape induces a tangential velocity that generates a swirl effect on the flow and hence, it promotes a heat transfer enhancement. 
    
    
    \item The corrugated geometry enhances the Nusselt number at the expense of a higher friction factor, averaging an increase for ${\rm Nu}$ and $f$ of around 8 and 6 times, respectively, when compared to the values for smooth tube.
    
    \item For low Reynolds numbers in the laminar regime, corrugated tubes are not very effective for heat transfer enhancement and may yield thermal performance factors $\eta$ smaller than unit. On the other hand, for high Reynolds numbers (${\rm Re} \approx 2300$), the values of $\eta$ were found to lay in the range of $1.5-2$ for all investigated corrugated tubes.
    
    \item The estimated thermal performance factor $\eta$ shows that the most suitable of the three investigated corrugated geometries is the one with the lower corrugation pitch (\textit{i.e.} higher severity index), being recommended to operate at the start of the turbulent regime ($1000 \lesssim {\rm Re} \lesssim 2300$) for yielding values of $\eta$ in the range of $3$ to $5$. In this respect, a HEX employing corrugated tubes should be designed considering such flow conditions.
   
\end{itemize}

Overall, one can conclude from the present study that the corrugated tubes perform considerably better from $700 \lesssim {\rm Re} \lesssim 2300$ as compared to the smooth tube. Among them, the tube with lowest pitch (P6~$=$~6~mm) has the highest thermal performance factor, and so it will be the best option for practical applications. As compared to the smooth tube, the corrugated tubes allow to attain a turbulent flow regime at low working fluid mass flow rate, which is a key feature for the optimization of the CFT-HEX for vehicle Rankine cycle waste heat recovery systems. Besides promoting an early laminar-turbulent transition, the tube helical corrugation induces a more severe swirl effect as the Reynolds number increases. Even so, this benefit is less significant for ${\rm Re} \gtrsim 2300$ when the flow inside the smooth tube becomes turbulent.

Regarding the future work, further reduce the critical Reynolds number for the start of transition is of crucial importance to further optimize the heat exchanger. The present study revealed that the critical Reynolds number for the start of transition is the same for tubes with some corrugation height and different helical pitch. Taking this into account, in a future research the influence of corrugation height on the critical Reynolds number for the start of transition will be investigated both experimentally and numerical.


\section*{Acknowledgements}
Authors are grateful to FCT - Fundacao para a Ci\^encia e a Tecnologia, I.P., for partially financing the research under the framework of the project nr 030171 financed by LISBOA-01-0145-FEDER-030171/PTDC/EME-SIS/30171/2017 and of the project JICAM/0003/2017. This work has been also supported by national funds through FCT under the project ’UIDB/50022/2020’.

\nomenclature[a]{$A_c$}{Cross-section area [m$^2$]}
\nomenclature[a]{$A_s$}{Surface area [m$^2$]}
\nomenclature[a]{$c_p$}{Fluid specific heat capacity [J/(kg.K)]}
\nomenclature[a]{$D_h$}{Hydraulic diameter [m]}
\nomenclature[a]{$D_i$}{Inner diameter [m]}
\nomenclature[a]{$e$}{Corrugation height [m]}
\nomenclature[a]{$f$}{Friction factor [-]}
\nomenclature[a]{$\rm Gz$}{Graetz number [-]} 
\nomenclature[a]{$h$}{Convective heat transfer coefficient [W/(m$^2$.K)]}
\nomenclature[a]{$k$}{Fluid thermal conductivity [W/(m.K)]} 
\nomenclature[a]{$k_w$}{Tube wall thermal conductivity [W/(m.K)]} 
\nomenclature[a]{$L$}{Tube length [m]}
\nomenclature[a]{$\Dot{m}$}{Mass flow rate [kg/s]}
\nomenclature[a]{$\rm Nu$}{Nusselt number [-]} 
\nomenclature[a]{$p$}{Corrugation pitch  [m]}
\nomenclature[a]{$\rm Pr$}{Prandtl number [-]}
\nomenclature[a]{$P_w$}{Wetted perimeter [m]}
\nomenclature[a]{$P$}{Pressure [Pa]}
\nomenclature[a]{$\Delta P$}{Pressure drop [Pa]}
\nomenclature[a]{$\Dot{Q}$}{Heat rate [W]}
\nomenclature[a]{$r$}{radial coordinate [m]}
\nomenclature[a]{$R$}{radius of tube wall [m]}
\nomenclature[a]{$\rm Re$}{Reynolds number [-]}
\nomenclature[a]{$\rm Ri$}{Richardson number [-]}
\nomenclature[a]{$S_m$}{swirl number}
\nomenclature[a]{$T$}{Temperature [ºC]}
\nomenclature[a]{$\Delta T$}{Temperature difference [ºC]}
\nomenclature[a]{$t$}{Wall thickness [m], time [s]}
\nomenclature[a]{$u$}{Velocity magnitude [m/s]}
\nomenclature[a]{$u_{axial}$}{Axial velocity component [m/s]}
\nomenclature[a]{$u_m$}{Average axial velocity [m/s]}
\nomenclature[a]{$u_{tang}$}{Tangential velocity component [m/s]}
\nomenclature[a]{$\bf v$}{Velocity vector [m/s]}
\nomenclature[a]{$x$}{Axial coordinate[m]}
\nomenclature[g]{$\rho$}{Fluid density [kg/m$^3$]}
\nomenclature[g]{$\mu$}{Fluid dynamic viscosity [kg/m.s]}
\nomenclature[g]{$\Phi$}{Viscous dissipation function [s$^{-2}$]}
\nomenclature[g]{$\phi$}{Severity index [-]}
\nomenclature[g]{$\eta$}{Thermal performance factor [-]}
\nomenclature[s]{$1$}{Test section}
\nomenclature[s]{$2$}{Last third of test section}
\nomenclature[s]{$cr$}{Critical}
\nomenclature[s]{$f$}{Fluid}
\nomenclature[s]{$in$}{Inlet}
\nomenclature[s]{$out$}{Outlet}
\nomenclature[s]{$ref$}{Reference}
\nomenclature[s]{$s,i$}{External tube surface, position $i$}
\nomenclature[s]{$w,i$}{Inner tube wall, position $i$}
\nomenclature[t]{$-$}{Average}
\nomenclature[t]{$+$}{Dimensionless}
\nomenclature[t]{$\sim$}{Modified (based on inner diameter)}

\printnomenclature

	%
	\bibliographystyle{unsrt}
	
	\bibliography{library}

\begin{thebibliography}{10}

\bibitem{Franco2002}
Alessandro Franco and Alessandro Russo.
\newblock {Combined cycle plant efficiency increase based on the optimization
  of the heat recovery steam generator operating parameters}.
\newblock {\em International Journal of Thermal Sciences}, 41(9):843--859,
  2002.

\bibitem{Wang2011}
Tianyou Wang, Yajun Zhang, Zhijun Peng, and Gequn Shu.
\newblock {A review of researches on thermal exhaust heat recovery with Rankine
  cycle}.
\newblock {\em Renewable and Sustainable Energy Reviews}, 15(6):2862--2871,
  2011.

\bibitem{Sheikholeslami2015}
Mohsen Sheikholeslami, Mofid Gorji-Bandpy, and Davood~Domiri Ganji.
\newblock {Review of heat transfer enhancement methods: Focus on passive
  methods using swirl flow devices}, sep 2015.

\bibitem{Ji2015}
Wen~Tao Ji, Anthony~M. Jacobi, Ya~Ling He, and Wen~Quan Tao.
\newblock {Summary and evaluation on single-phase heat transfer enhancement
  techniques of liquid laminar and turbulent pipe flow}.
\newblock {\em International Journal of Heat and Mass Transfer}, 88:735--754,
  2015.

\bibitem{WEI2021100948}
Haoran Wei, Hazim Moria, Kottakkaran~Sooppy Nisar, Raymond Ghandour, Alibek
  Issakhov, Yu-Liang Sun, Amr Kaood, and Mohammad~Mehdizadeh Youshanlouei.
\newblock Effect of volume fraction and size of al2o3 nanoparticles in thermal,
  frictional and economic performance of circumferential corrugated helical
  tube.
\newblock {\em Case Studies in Thermal Engineering}, 25:100948, 2021.

\bibitem{WANG201951}
Guiqing Wang, Cong Qi, Maoni Liu, Chunyang Li, Yuying Yan, and Lin Liang.
\newblock Effect of corrugation pitch on thermo-hydraulic performance of
  nanofluids in corrugated tubes of heat exchanger system based on exergy
  efficiency.
\newblock {\em Energy Conversion and Management}, 186:51--65, 2019.

\bibitem{Garcia2012}
A.~Garc{\'{i}}a, J.~P. Solano, P.~G. Vicente, and A.~Viedma.
\newblock {The influence of artificial roughness shape on heat transfer
  enhancement: Corrugated tubes, dimpled tubes and wire coils}.
\newblock {\em Applied Thermal Engineering}, 35(1):196--201, 2012.

\bibitem{Kareem2015}
Zaid~S Kareem, M~N {Mohd Jaafar}, Tholudin~M Lazim, Shahrir Abdullah, and
  Ammar~F Abdulwahid.
\newblock {Passive heat transfer enhancement review in corrugation}.
\newblock {\em Experimental Thermal and Fluid Science}, 68:22--38, 2015.

\bibitem{QIAN2020119876}
Jin yuan Qian, Chen Yang, Min rui Chen, and Zhi jiang Jin.
\newblock Thermohydraulic performance evaluation of multi-start spirally
  corrugated tubes.
\newblock {\em International Journal of Heat and Mass Transfer}, 156:119876,
  2020.

\bibitem{AKBARZADEH2020735}
Sanaz Akbarzadeh and Mohammad~Sadegh Valipour.
\newblock Energy and exergy analysis of a parabolic trough collector using
  helically corrugated absorber tube.
\newblock {\em Renewable Energy}, 155:735--747, 2020.

\bibitem{AKBARZADEH2021101013}
Sanaz Akbarzadeh and Mohammad~Sadegh Valipour.
\newblock The thermo-hydraulic performance of a parabolic trough collector with
  helically corrugated tube.
\newblock {\em Sustainable Energy Technologies and Assessments}, 44:101013,
  2021.

\bibitem{Liu2013}
Lin Liu, Xiang Ling, and Hao Peng.
\newblock {Analysis on flow and heat transfer characteristics of EGR helical
  baffled cooler with spiral corrugated tubes}.
\newblock {\em Experimental Thermal and Fluid Science}, 44:275--284, 2013.

\bibitem{Wang2018b}
Wei Wang, Yaning Zhang, Yongji Li, Huaizhi Han, and Bingxi Li.
\newblock {Multi-objective optimization of turbulent heat transfer flow in
  novel outward helically corrugated tubes}.
\newblock {\em Applied Thermal Engineering}, 138:795--806, jun 2018.

\bibitem{Meyer_2014}
Josua~P. Meyer.
\newblock Heat transfer in tubes in the transitional flow regime.
\newblock In {\em Proceedings of the 15th International Heat Transfer
  Conference}. Begellhouse, 2014.

\bibitem{ANDRADE2019940}
F.~Andrade, A.S. Moita, A.~Nikulin, A.L.N. Moreira, and H.~Santos.
\newblock Experimental investigation on heat transfer and pressure drop of
  internal flow in corrugated tubes.
\newblock {\em International Journal of Heat and Mass Transfer}, 140:940 --
  955, 2019.

\bibitem{Mohammed2013}
H.~A. Mohammed, Abdalrazzaq~K. Abbas, and J.~M. Sheriff.
\newblock {Influence of geometrical parameters and forced convective heat
  transfer in transversely corrugated circular tubes}.
\newblock {\em International Communications in Heat and Mass Transfer},
  44:116--126, 2013.

\bibitem{CORCOLES2020106526}
J.I. Córcoles, J.D. Moya-Rico, A.E. Molina, and J.A. Almendros-Ibáñez.
\newblock Numerical and experimental study of the heat transfer process in a
  double pipe heat exchanger with inner corrugated tubes.
\newblock {\em International Journal of Thermal Sciences}, 158:106526, 2020.

\bibitem{YANG2020106520}
Chen Yang, Gan Liu, Junhui Zhang, and Jin yuan Qian.
\newblock Thermohydraulic analysis of hybrid smooth and spirally corrugated
  tubes.
\newblock {\em International Journal of Thermal Sciences}, 158:106520, 2020.

\bibitem{Vicente2004}
P.G. Vicente, A.~Garcia, and A.~Viedma.
\newblock {Mixed convection heat transfer and isothermal pressure drop in
  corrugated tubes for laminar and transition flow}.
\newblock {\em International Communications in Heat and Mass Transfer},
  31(5):651--662, jul 2004.

\bibitem{Vicente2004a}
P.~G. Vicente, A.~Garcia, and A.~Viedma.
\newblock {Experimental investigation on heat transfer and frictional
  characteristics of spirally corrugated tubes in turbulent flow at different
  Prandtl numbers}.
\newblock {\em International Journal of Heat and Mass Transfer},
  47(4):671--681, feb 2004.

\bibitem{Dizaji2016}
Hamed~Sadighi Dizaji and Samad Jafarmadar.
\newblock {Experiments on New Arrangements of Convex and Concave Corrugated
  Tubes through a Double-pipe Heat Exchanger}.
\newblock {\em Experimental Heat Transfer}, 29(5):577--592, 2016.

\bibitem{Pethkool2011}
S.~Pethkool, S.~Eiamsa-ard, S.~Kwankaomeng, and P.~Promvonge.
\newblock {Turbulent heat transfer enhancement in a heat exchanger using
  helically corrugated tube}.
\newblock {\em International Communications in Heat and Mass Transfer},
  38(3):340--347, 2011.

\bibitem{Bergles_Blumenkrantz_Taborek_1974}
A.~E. Bergles, A.~R. Blumenkrantz, and Jerry Taborek.
\newblock Performance evaluation criteria for enhanced heat transfer surfaces.
\newblock In {\em Proceeding of International Heat Transfer Conference 5}.
  Begellhouse, 1974.

\bibitem{ZIMPAROV19941807}
Ventsislav~D. Zimparov and Nikolai~L. Vulchanov.
\newblock Performance evaluation criteria for enhanced heat transfer surfaces.
\newblock {\em International Journal of Heat and Mass Transfer}, 37(12):1807 --
  1816, 1994.

\bibitem{Wang2018a}
Wei Wang, Yaning Zhang, Yongji Li, Huaizhi Han, and Bingxi Li.
\newblock {Numerical study on fully-developed turbulent flow and heat transfer
  in inward corrugated tubes with double-objective optimization}.
\newblock {\em International Journal of Heat and Mass Transfer}, 120:782--792,
  may 2018.

\bibitem{bejan2013convection}
A.~Bejan.
\newblock {\em Convection Heat Transfer}.
\newblock John Wiley \& Sons, Ltd, 2013.

\bibitem{Corcoles-Tendero2018a}
J.~I. C{\'{o}}rcoles-Tendero, J.~F. Belmonte, A.~E. Molina, and J.~A.
  Almendros-Ib{\'{a}}{\~{n}}ez.
\newblock {Numerical simulation of the heat transfer process in a corrugated
  tube}.
\newblock {\em International Journal of Thermal Sciences}, 126:125--136, apr
  2018.

\bibitem{RABIENATAJDARZI2021106759}
A.~Ali {Rabienataj Darzi}, Mohamad Abuzadeh, and Mohamad Omidi.
\newblock Numerical investigation on thermal performance of coiled tube with
  helical corrugated wall.
\newblock {\em International Journal of Thermal Sciences}, 161:106759, 2021.

\bibitem{Nikulin2019}
A.~Nikulin, A.~S. Moita, A.~L.N. Moreira, S.~M.S. Murshed, A.~Huminic,
  Y.~Grosu, A.~Faik, J.~Nieto-Maestre, and O.~Khliyeva.
\newblock {Effect of Al2O3 nanoparticles on laminar, transient and turbulent
  flow of isopropyl alcohol}.
\newblock {\em International Journal of Heat and Mass Transfer},
  130:1032--1044, 2019.

\bibitem{InternationalHandbookCommittee1997}
ASM {International Handbook Committee}.
\newblock {Metals Handbook Volume 20 - Materials Selection and Design}.
\newblock {\em ASM International: Materials Park, OH}, 1997.

\bibitem{Taylor1994}
Barry~N Taylor and Chris~E Kuyatt.
\newblock {Guidelines for Evaluating and Expressing the Uncertainty of NIST
  Measurement Results}.
\newblock {\em NIST Technical Note}, 1994.

\bibitem{star}
{STAR-CCM+}.
\newblock
  \url{https://www.plm.automation.siemens.com/global/en/products/simcenter/STAR-CCM.html}.

\bibitem{solid}
{3D CAD Design Software | SOLIDWORKS}.
\newblock \url{https://www.solidworks.com/}.

\bibitem{Patankar1980}
S.~V. Patankar.
\newblock {\em {Numerical heat transfer and fluid flow.}}
\newblock CRC Press, 1980.

\bibitem{Haervig2017}
J.~H{\ae}rvig, K.~S{\o}rensen, and T.~J. Condra.
\newblock {On the fully-developed heat transfer enhancing flow field in
  sinusoidally, spirally corrugated tubes using computational fluid dynamics}.
\newblock {\em International Journal of Heat and Mass Transfer},
  106:1051--1062, 2017.

\bibitem{XIN2018402}
Feng Xin, Zhichun Liu, Nianben Zheng, Peng Liu, and Wei Liu.
\newblock Numerical study on flow characteristics and heat transfer enhancement
  of oscillatory flow in a spirally corrugated tube.
\newblock {\em International Journal of Heat and Mass Transfer}, 127:402--413,
  2018.

\bibitem{JORGE2019397}
Pedro Jorge, Miguel~A.A. Mendes, Eric Werzner, and José~M.C. Pereira.
\newblock Characterization of laminar flow in periodic open-cell porous
  structures.
\newblock {\em Chemical Engineering Science}, 201:397 -- 412, 2019.

\bibitem{Roache1997}
P.~J. Roache.
\newblock {QUANTIFICATION OF UNCERTAINTY IN COMPUTATIONAL FLUID DYNAMICS}.
\newblock {\em Annual Review of Fluid Mechanics}, 1997.

\bibitem{bejan2003heat}
Adrian Bejan.
\newblock {\em Heat transfer handbook}.
\newblock J. Wiley, New York, 2003.

\bibitem{Hrycak1974}
P.~Hrycak and R.~Andrushkiw.
\newblock {CALCULATION OF CRITICAL REYNOLDS NUMBERS IN ROUND PIPES AND INFINITE
  CHANNELS AND HEAT TRANSFER IN TRANSITION REGIONS.}
\newblock 1974.

\bibitem{Petukhov1970}
B.~S. Petukhov.
\newblock {Heat Transfer and Friction in Turbulent Pipe Flow with Variable
  Physical Properties}.
\newblock {\em Advances in Heat Transfer}, 1970.

\bibitem{MEYER20111587}
J.P. Meyer and J.A. Olivier.
\newblock Transitional flow inside enhanced tubes for fully developed and
  developing flow with different types of inlet disturbances: Part i –
  adiabatic pressure drops.
\newblock {\em International Journal of Heat and Mass Transfer}, 54(7):1587 --
  1597, 2011.

\end{thebibliography}
\end{document}